\documentclass[graybox, nosecnum]{svmult}

\usepackage{mathptmx}       % selects Times Roman as basic font
\usepackage{helvet}         % selects Helvetica as sans-serif font
\usepackage{courier}        % selects Courier as typewriter font
\usepackage{type1cm}        % activate if the above 3 fonts are
\usepackage{makeidx}         % allows index generation
\usepackage{graphicx}        % standard LaTeX graphics tool
\usepackage{multicol}        % used for the two-column index
\usepackage[bottom]{footmisc}% places footnotes at page bottom
\usepackage{hyperref}        % for hyperlinks
\usepackage{soul}            % for high-lighting of text
\usepackage{ulem}            % just for comments, to be evicted in the final version
\usepackage{comment}         % just for comments, to be evicted in the final version

\hypersetup{colorlinks=true,urlcolor=blue}
\usepackage[square,numbers]{natbib}
\makeindex             % used for the subject index
\usepackage[utf8]{inputenc}
\usepackage{bm}
\usepackage{amssymb, amsmath}

\newcommand{\bea}{\begin{eqnarray}}
\newcommand{\eea}{\end{eqnarray}}
\newcommand{\beq}{\begin{equation}}
\newcommand{\eeq}{\end{equation}}

\begin{document}
\normalem %This is necessary to recover the standard behaviour of \emph{} under ulem package.
%ulem is necessary for ``strike out" in comments

% This "\tableofcontents{}" is used
% only to see the whole structure of the manuscript
% and will be commented out for submission.
%\textcolor{black}{}
%\tableofcontents{}
\title*{Post-Newtonian templates for gravitational waves from compact binary inspirals}
\author{Soichiro Isoyama \thanks{corresponding author},
Riccardo Sturani and Hiroyuki Nakano}
\institute{Soichiro Isoyama \at School of Mathematics,
University of Southampton, Southampton, United Kingdom. \email{isoyama@yukawa.kyoto-u.ac.jp}
\and Riccardo Sturani \at International Institute of Physics,
Universidade Federal do Rio Grande do Norte, Natal, Brazil. \email{riccardo.sturani@ufrn.br}
\and Hiroyuki Nakano \at Faculty of Law, Ryukoku University, Kyoto, Japan.
\email{hinakano@law.ryukoku.ac.jp}}
\maketitle

\abstract{To enable detection and maximise the physics output of gravitational wave observations from compact binary systems, it is crucial the availability of accurate waveform models. The present work aims at giving an overview for non-experts of the (inspiral) waveforms used in the gravitational wave data analysis for compact binary coalescence. We first provide the essential elements of gravitational radiation physics within a simple Newtonian orbital dynamics and the linearized gravity theory, describing the adiabatic approximation applied to binary systems: the key element to construct the theoretical gravitational waveforms in practice. We next lay out the gravitational waveforms in the post-Newtonian approximation to General Relativity, and highlight the basic input for the inspiral waveform of the slowly evolving, spinning, nonprecessing, quasicircular binary black holes, namely, post-Newtonian energy, fluxes and the (absorption-corrected) balance equation. The post-Newtonian inspiral templates are then presented both in the time and frequency domain. Finally, including the merger and subsequent ringdown phase, we briefly survey the two families of the full waveform models of compact binary mergers currently implemented in LSC Algorithm Library Simulation: the effective-one-body approach and the phenomenological frequency domain model.}

%%%%%%%%%%%%%%%%%%%%%%%%%%%%%%%%%%%%%%%%%%%%%%%%%%
\section{Keywords} 
%%%%%%%%%%%%%%%%%%%%%%%%%%%%%%%%%%%%%%%%%%%%%%%%%%

%\textcolor{red}{(Please provide keywords required to facilitate search of chapter on web;
%maximum 10 keywords.)}

Binary black holes, Compact object binaries,
Gravitational waves, Post-Newtonian approximation, 
Adiabatic approximation, 
Stationary phase approximation,
Inspiral-merger-ringdown waveforms,
LSC Algorithm Library.

%%%%%%%%%%%%%%%%%%%%%%%%%%%%%%%%%%%
%\section{\textit{Main Text}}
%%%%%%%%%%%%%%%%%%%%%%%%%%%%%%%%%%%

%%%%%%%%%%%%%%%%%%%%%%%%%%%%%%%%%%%%%%%%%%%%%%%%%%%%%%%%%%%%%%%%%%%%%%%%%
%This is the main body of the chapter and should include sections and subsections. 
%Please re-name this heading and add your own subheadings.\\
%{\bf Note: Footnotes should not be used! Avoid acknowledgements other than those related to funding.}

%%%%%%%%%%%%%%%%%%%%%%%%%%%%%%%%%%%%%%%%%%%%%%%%%%
\section{Introduction}
%%%%%%%%%%%%%%%%%%%%%%%%%%%%%%%%%%%%%%%%%%%%%%%%%%

%\textcolor{red}{(Introduction to the chapter; length depends on the topic describing importance of %subject and content)}

%%\rs{\sout{TO DO: add some PRL-style introduction for general audience, like PhDs in Kyoto.}}

The recent detection of gravitational waves (GW) from compact binary
coalescences made by the large interferometers LIGO~\cite{TheLIGOScientific:2014jea} 
and Virgo~\cite{TheVirgo:2014hva} opened the era of GW astronomy, 
triggering scientific interests over all aspects of GW production and detection.
This new, exciting branch of scientific activity is even bursting now, 
with the third GW interferometer: KAGRA online~\cite{Akutsu:2020his}; 
for a useful summary of these detectors see Chapter 2 by Grote and Dooley 
in this book.

Because the GW signals are in general drowned in a much larger noise, their extraction from data stream
and correct interpretation crucially depend on the quality of theoretically predicted template waveforms.
Experimental data are processed via matched-filtering techniques; see, e.g., Ref.~\cite{Allen:2005fk},
which are particularly sensitive to the \emph{phase} of GW signals; hence, a prediction of waveforms
with absolute $\ll {\cal O}(1)$ precision (in the phase) is very important, 
especially for a correct parameter estimation and 
in general for maximising the physics output of detection 
(see, e.g., Chapter~43 by Krolak in this book for differences between GW searches and parameter estimations).

From the point of view of theoretical modelling, 
the process of binary coalescence can be divided into three distinct phases:
inspiral, merger and ringdown.
During the inspiral phase, due to large binary separation with respect to 
their size, the dynamics can be efficiently described 
within the post-Newtonian (PN) approximation to general relativity (GR),
the small expansion parameter being 
\bea
\frac{G M}{c^2 R} \sim \frac{v^2}{c^2} \ll 1 \,,
\label{PNparam0}
\eea
where $M$, $v$, and $R$ denote, respectively, the total mass, 
characteristic orbital velocity, and characteristic separation of the binary
system. A $n$-PN correction to the gravitational potential is then given
by any term of the type $(M/R)^{n-j}v^{2j}$ with integers $0\leq j\leq n$.
The GR corrections to the Newtonian dynamics around flat Minkowski spacetime 
are then incorporated into the equation of motion and Einstein's field equations
(expanded around the Minkowski metric) order by order in that small parameter. 
An excellent textbook devoted to the PN theory is \textit{Gravity} by 
Poisson and Will~\cite{2014grav.book.....P}; more expert-oriented reviews 
(including more recent developments) 
are provided by Blanchet ~\cite{Blanchet:2013haa}, 
Sch\"{a}fer and Jaranowski~\cite{Schafer:2018kuf}, 
and Futamase and Itoh~\cite{Futamase:2007zz}.
Recently, the PN approximation to GR has been developed in an effective field theory description~\cite{Goldberger:2004jt}, dubbed non-relativistic GR (NRGR), 
which resulted in an alternative,
independent and equivalent derivation of the PN dynamics 
(see Refs.~\cite{Goldberger:2007hy,Foffa:2013qca,Porto:2016pyg,Levi:2018nxp} for reviews, 
and Chapter~32 by Sturani in this book).

It is a remarkable feature of GR that the two-body dynamics in the inspiral phase
is exactly the one of two (structureless) point particles 
up to $5$PN when finite finite-size effects come into play, 
according to the \emph{effacement principle}~\cite{Damour1984}.
However, finite size effects 
that tidally ``deform'' a black hole (BH)
have been shown to vanish for a spherical, non-spinning BHs
~\cite{Binnington:2009bb,Damour:2009va,Kol:2011vg,Gurlebeck:2015xpa} in the static limit, 
and there are indications that they vanish 
for spinning BHs, too~\cite{Poisson:2014gka,Pani:2015nua} 
(see also~\cite{LeTiec:2020spy,Chia:2020yla,LeTiec:2020spyx}).
On the other hand the tidal deformation effects for \emph{material} bodies 
like neutron stars (NS) are expected, depending on the equation of state of matter in NSs,  
and there have been some evidence of their detection in GW170817~\cite{TheLIGOScientific:2017qsa}.

Considering circular orbit, in the nonprecessing case, is usually accurate 
since angular momentum radiation is more efficient than energy radiation;
thus, binary orbits tend to circularize at early stage~\cite{Peters:1964zz}. 
However, while unlikely, it is not excluded that some observed GW signals 
can show a non-negligible eccentricity
due to, e.g., dynamical formation of binary BHs (BBHs)
presented in Chapter~15 by Kocsis in this book.
When spins are included in the binary dynamics, in general, 
one has to also take into account orbital precession caused 
by the orbital angular momentum-spin coupling, as well as the relativistic precession of the spins.

As the binary's orbital separation shrinks by decreasing the binding energy
through the emission of GWs (with an increase in relative velocity of the constituents), 
the system enters the merger phase where two compact objects
form one single object.
At the merger phase the PN approximation breaks down, $v \lesssim 1$, 
and full numerical-relativistic simulations have to be used, 
where the nonlinear radiative dynamics of the binaries is obtained 
by directly solving the exact Einstein field equations numerically: 
see, e.g., a recent text by Baumgarte and Shapiro~\cite{Baumgarte:2010ndz} 
and Chapter~34 by Zhao et al. in this book.
Immediately after the merger a perturbed BH forms
which rapidly damps its excitation in a {ringdown} phase, whose
GW emission is described by a superposition of damped sinusoids~\cite{Nollert:1999ji,Kokkotas:1999bd,Berti:2009kk}; 
the detail can be found in Chapter~38 by Konoplya in this book. 

\begin{figure}[th]
    \centering
    \includegraphics[width=0.8\textwidth]{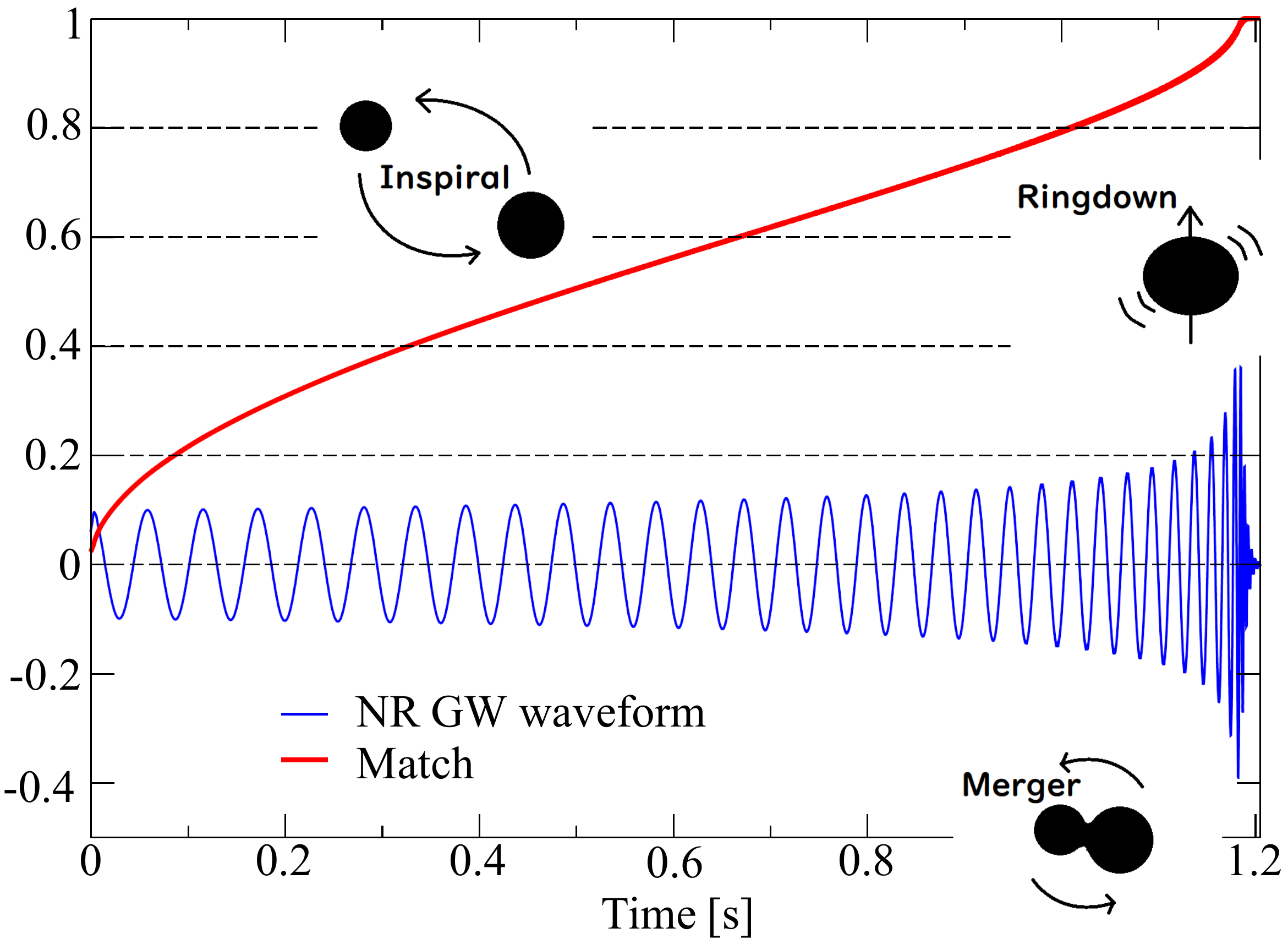}
    \caption{
    Inspiral-merger-ringdown (IMR) signal 
    from the numerical-relativity simulation SXS:BBH:0305~\cite{Boyle:2019kee} which models GW150914.
    The horizontal axis is the real time.
    Here, we present only the (scaled) ``$+$'' GW mode
    (blue thin curve). The red thick curve shows the ``match'', i.e. the
    accumulation of the (normalized) signal-to-noise ratio in white noise.}
    \label{fig:IMR}
\end{figure}

Current inspiral-merger-ringdown (IMR) waveform models for data analysis applications 
combine input from the PN theory and NR simulations.  
The models implemented in the LSC Algorithm Library (LAL) Simulation~\cite{lalsuite} 
are summarized in ``Enumeration Type Documentation'' in Ref.~\cite{lalsim}.
The GW signal models used for the analyses of observed GW events so far are described in: 
\begin{description}
\item GW150914:
  first BBH detection~\cite{Abbott:2016blz,TheLIGOScientific:2016wfe},
\item GW170817:
first binary NS (BNS) detection~\cite{TheLIGOScientific:2017qsa,Abbott:2018wiz},
\item events from the first Gravitational Wave Transient Catalog:
  first and second observation runs, O1 and O2 henceforth, including the above
  two GW events, along with
  GW151012, GW151226, GW170104, GW170608, GW170729, GW170809, GW170814, GW170818, GW170823 ~\cite{LIGOScientific:2018mvr},
\item GW190412:
  O3, unequal mass (mass ratio $\sim 3$ with nonzero spin for the heavier object)
  BBH~\cite{LIGOScientific:2020stg},
\item GW190425:
O3, second BNS detection~\cite{Abbott:2020uma},
\item GW190814:
  O3, highest mass ratio between binary constituents ($\sim 10$) and
  lighter object in the low ``mass gap'' ~\cite{Abbott:2020khf},
\item GW190521:
  O3, whose remnant can be considered as the first ever evidence of a light
  intermediate mass ($\sim 10^2 M_\odot$) BH ~\cite{Abbott:2020tfl,Abbott:2020mjq},
\item events from the second Gravitational Wave Transient Catalog:
  the first half of the third observing run (O3a), including 39 GW events~\cite{Abbott:2020niy}.
\end{description}

%%For notable events in O3, the IMR waveform models for the parameter estimation 
%%of the binaries are shown:
%%\begin{description}
%%\item GW190412:
%%\url{https://doi.org/10.7935/20yv-ka61}
%%\item GW190425:
%%\url{https://doi.org/10.7935/ggb8-1v94}
%%\item GW190814:
%%\url{https://doi.org/10.7935/zzw5-ak90}
%%\item GW190521:
%%\url{https://doi.org/10.7935/1502-wj52} 
%%\end{description}
%%in ``Gravitational Wave Open Science Center''
%%at \url{https://www.gw-openscience.org/about/}.%~\cite{GWOSC}

%end intro

%%%%%%%%%%%%%%%%%%%%%%%%%%%%%%%%%%%%%%%%%%%%%%%%%%%%%%%%%%%
\subsection{Goal and relation to other chapters}
%%%%%%%%%%%%%%%%%%%%%%%%%%%%%%%%%%%%%%%%%%%%%%%%%%%%%%%%%%%%%%%

Our purpose with this chapter is twofold: 
(i) to make a compilation of the PN-based GW
  waveform models for the binary inspirals in literature;
(ii) to provide a ``catalogue'' of IMR waveform families 
implemented in the LALSimulation.
The main target audience is graduate students (as well as researchers) 
who work in areas outside of the GW source modelling of binaries,  
but who need a working knowledge of waveform models.
%
%\textcolor{red}{(Include a list of related entries from the handbook here 
%that may be of further interest to the readers. e.g. Riccardo's EFT review.)}
%
%\sis{TO DO: updated this after finalising overall structure:} 
%

We set the stage with the basics of gravitational radiation physics,  
using the simple quadrupole formalism in the linearized gravity theory. 
We next cover the GW waveforms in the PN theory, 
collecting the state of art of PN binding energy and energy flux 
of slowly evolving, spinning, and nonprecessing BBHs, 
and present their inspiral PN templates both in the time and frequency domains.
Then, we briefly overview the two main families of the full IMR waveforms: 
the effective-one-body approach and the phenomenological frequency-domain model, 
used and implemented in LALSimulation so far. 
We conclude our chapter with some of remaining issues and future prospects.
%%%%%%%%%%%%%%%%%%%%%%%%%%%%%%%%%%%%%%%%%%%%%%%%%%%%%%%%%%%%%%%%%%%%%%%%%%%%%%%%%%%%%%%%%%%

%%%%%%%%%%%%%%%%%%%%%%%%%%%%%%
\subsection{Notations}
We use the metric signature $(-+++)$ and the standard geometrized unit $(c = 1 = G)$ 
with the useful conversion factor $1 M_{\odot} = 1.477 \; {\rm{km}} 
= 4.926 \times 10^{-6} \; {\rm{s}}$, throughout.
Greek indices $\mu,\,\nu,\,\dots$ denote $0,\,1,\,2,\,3$ (or, e.g., $t,\,x,\,y,\,z$)  
while the Latin letters $i,\,j,\,\dots$ run over $1,\,2,\,3$.
%which means that the flat-space Minkowski metric is defined by diag$(-1,\,1,\,1,\,1)$. 

For a binary with masses $m_a$ and spin vector ${\bf S}_a$ 
(where $a=1,\,2$ labels each compact object), 
we will use the notation defined in Ref.~\cite{LIGOScientific:2018mvr}. 
The total mass of a binary is
\beq
M \equiv m_1 + m_2 \,,
\eeq
and the mass ratio is 
%given by
\beq
q \equiv \frac{m_2}{m_1} \leq 1 \,.
\eeq
We also use the symmetric mass ratio 
\beq
\eta \equiv \frac{m_1 m_2}{M^2} \,,
\eeq
and the reduced mass $\mu \equiv \eta \,M$.
The spin vector ${\bf S}_a$ has a conserved magnitude, 
as long as absorption effects are neglected, 
by using an appropriate spin supplementary condition 
to remove unnecessary degrees of freedom associated with the spin, 
and we denote its dimensionless magnitude by 
\beq
{\chi_a} \equiv |\bm{\chi}_a| \equiv \frac{|{{\bf S}_a|}}{m_a^2}\,.
\eeq
Finally, the $O(v^{2n})$ terms relative to the Newtonian dynamics 
will be referred to as the $n$-th PN order.

%%%%%%%%%%%%%%%%%%%%%%%%%%%%%%%%%%%%%%%%%%%%%%%%%%%%%%%%%%%%%%%%%%%%%%%%%%%
\section{The essence: Quadrupole radiation from a mass in circular orbit}
%%%%%%%%%%%%%%%%%%%%%%%%%%%%%%%%%%%%%%%%%%%%%%%%%%%%%%%%%%%%%%%%%%%%%%%%%%%%

The essence of GW generation formalism for a binary inspiral can be 
understood from the Newtonian orbital dynamics and the Einstein's quadrupole formalism 
based on the linearized gravity theory.
While a raw approximation, the linearized gravity theory catches
the basic concepts behind the GW signal calculation, 
without the intricacies of the full GR nonlinearity
(see, e.g., Chapter~11 of Poisson and Will~\cite{2014grav.book.....P}
for rigorous derivation of the quadrupole formula and details about
the PN treatment).
We here review the main line of reasoning and results, modelled on nice
tutorials by Flanagan and Hughes~\cite{Flanagan:2005yc} and the LVC collaboration~\cite{Abbott:2016bqf}. 
The material covered in this section is fairly standard, 
and the details of derivation are given in various introductory GW and GR texts, 
including Landau and Lifshitz~\cite{Landau:1982dva} and Maggiore~\cite{Maggiore:1900zz}. 

We start by supposing that the full spacetime metric (i.e., gravitational field) $g_{\mu \nu}$ 
deviates only slightly from the background flat metric 
$\eta_{\mu \nu} \equiv {\rm {diag}}(-1,\,1,\,1,\,1)$: 
\beq
g_{\mu \nu} = \eta_{\mu \nu} + h_{\mu \nu},\,
\quad
||h_{\mu \nu}|| \ll 1\,,
\label{def-h}
\eeq
where $h_{\mu \nu}$ is a small metric perturbation (i.e., weak gravitational field) of 
the background Minkowski spacetime, and $||h_{\mu \nu}||$ is a typical magnitude 
of $h_{\mu \nu}$.

In the linearized gravity theory, everything is consistently expanded 
to linear order in $h_{\mu \nu}$, neglecting all higher-order terms, 
and all indices are raised and lowered with the Minkowski metric $\eta_{\mu \nu}$. 
Also, we assume that the decomposition of the metric~\eqref{def-h} is always preserved 
in any coordinate system that we can chose. 
The general covariance of full GR 
is then restricted to an infinitesimal coordinate transformation 
\beq
{x'}^{\mu} = x^{\mu} + \xi^{\mu} \,,
\eeq
where $\xi^{\mu}(x)$ is an infinitesimal vector field, 
and the metric perturbation changes via 
\beq
{h'}_{\mu \nu} 
= h_{\mu \nu} - \partial_{\mu} \xi_{\nu} - \partial_{\nu} \xi_{\mu}\,.
\label{gauge-h}
\eeq
(Note the close analogy to the gauge transformation of the vector potential $A^{\mu}$ 
in electromagnetic theory, i.e., ${A'}_{\mu} = A_{\mu} - \partial_{\mu} \chi$ 
with a scalar field $\chi$. This is why Eq.~\eqref{gauge-h} is often referred to as 
a gauge transformation in the linearized gravity theory). 

The Einstein's field equations in the linearized theory are best described 
under the Lorenz-gauge conditions 
\beq
\partial^{\mu} {\bar h}_{\mu \nu} = 0\,,
\label{Lgauge}
\eeq
in terms of the trace-reversed perturbation 
\beq
{\bar h}_{\mu \nu} \equiv h_{\mu \nu} 
- \frac{1}{2} \eta_{\mu \nu} h^{\rho}{}_{\rho} \,,
\eeq
which can be enforced without loss of generality by making use of the coordinate freedom~\eqref{gauge-h}.
The Lorenz-gauge conditions reduce the Einstein's equations to a simple, decoupled wave equation for ${\bar h}_{\mu \nu}$ given by 
\beq
\Box \, {\bar h}_{\mu \nu} = -16 \pi T_{\mu \nu}\,, 
\label{linearized-Eeq}
\eeq
where $\Box \equiv \partial^{\mu} \partial_{\mu}$ 
is the d'Alembertian operator, 
and $T_{\mu \nu}$ is the energy-momentum tensor of the matter. 
Equation~\eqref{linearized-Eeq} can be solved by the method of Green's function $G(x,\,x')$ 
(for the operator $\Box$), imposing suitable boundary conditions. 
For the problem of the GW radiation, just like in electromagnetism, 
the appropriate choice is the retarded Green's function: 
\beq
G({x},\,{x}') = -\frac 1{4\pi} \frac{\delta (t' - t_{\rm ret})}{|{\bf x} - {\bf x}'|}\,,
\label{retG}
\eeq
where $t_{\rm ret} \equiv t - |{\bf x} - {\bf x}'|$ is the retarded time, 
accounting for the propagation delay between the source
at ${\mathbf x'}$ and the observer at ${\mathbf x}$. 
The associated retarded solution of Eq.~\eqref{linearized-Eeq} is then given by 
\beq
{\bar h}_{\mu \nu} 
= 
4\int d^3 x' \, 
\frac{T_{\mu \nu} (t - |{\bf x} - {\bf x}'| ,\, {\bf x}')}{ |{\bf x} - {\bf x}'|}\,.
\label{soln0_barh}
\eeq

For the GW calculation, again like the radiation in the electromagnetism, 
we are particularly interested in the behavior of the solution~\eqref{linearized-Eeq} 
in the (far-away) wave zone, where the length of the position vector $r \equiv |{\bf x}|$ 
is larger than the characteristic wavelength of the GW radiation $\lambda_{\rm Char}$ 
(defined by $\lambda_{\rm Char} = t_{\rm Char}$ in terms of the characteristic time scale of the 
change in the source $t_{\rm Char}$).
The characteristic radius $r_{\rm Char}\sim v \,t_{\rm Char}$ of the matter distribution is then 
smaller than $\lambda_{\rm Char}$, justifying the expansion
$ |{\bf x} - {\bf x}'| = r - n^{i} {x'}_{i} + O(r_{\rm Char}^2 / r)$ with $n^{i} \equiv x^{i} / r$, 
the zero-th solution in the (far-away) wave zone is given by
\beq
{\bar h}_{\mu \nu} = \frac{4}{r} \int d^3 x' \, {T_{\mu \nu} (t - r ,\, {\bf x}')}\,.
\label{soln1_barh}
\eeq
Here, not only the terms that fall off as $O(1/r^2)$ and higher are neglected,
but there is an additional approximation: all elements of the extended source
contribute to field at the same retarded time.

At this point, we must identify the truly radiative degrees of freedom contained 
in Eq.~\eqref{soln1_barh}. 
Apparently, all six degrees of freedom in ${h}_{\mu \nu}$
(${h}_{\mu \nu}$ have ten components, and four are constrained 
by the Lorenz-gauge condition~\eqref{Lgauge}) look radiative. 
However, they are not; only two are actually radiative. 
The remaining four are nonradiative degrees of freedom tied to the matter  
(like the Coulomb piece of the electromagnetic field), 
and their wave-like behaviours are merely artefact 
due to the Lorenz-gauge formulation of linealized Einstein equation~\eqref{linearized-Eeq}.
Indeed, a detailed analysis of the linerlized gravity theory 
by Flanagan and Hughes~\cite{Flanagan:2005yc} reveals that 
the radiative degrees of freedom are just encoded 
in the spatial transverse-traceless components $h_{i j}^{\rm {TT}}$ of ${h}_{\mu \nu}$, 
which satisfies the following four conditions (in addition to the Lorenz gauge condition): 
\beq
\partial^{i} h_{i j}^{\rm {TT}} = 0,\,
\quad
h^{i \,{\rm TT}}_{~i} = 0\,.
\label{TT}
\eeq
Importantly, $h_{i j}^{\rm {TT}}$ is coordinate-invariant (gauge-invariant) 
under the transformation~\eqref{gauge-h} (as long as the full metric~\eqref{def-h} 
remains asymptotically flat), and hence, it is the direct observable 
of GW detectors, e.g., LIGO, Virgo, and KAGRA. 
In the (far-away) wave zone, $h_{i j}^{\rm {TT}}$ can be obtained by projecting 
${h}_{\mu \nu}$ onto the plane orthogonal to the direction of propagation ${\bf n} = {\bf x} / r$ 
and subtracting its trace. We introduce the projector 
(this is sometimes referred to as Lambda tensor~\cite{Maggiore:1900zz}) 
\beq
\Lambda_{ij,\,kl}  \equiv P_{ik} P_{j l} - \frac{1}{2} P_{ij} P_{kl}\,,
\label{def-Lambda}
\eeq
where $P_{i j} = \delta_{i j} - n_i n_j$, and we obtain 
\beq
{\bar h}_{i j}^{\rm TT} = \frac{4}{r}\, \Lambda_{ij,\,kl}\, 
\int d^3 x' \, T^{k l}  (t - r ,\, {\bf x}')\,.
\label{barh0_TT}
\eeq

For the final step, we simplify the integral on the right of Eq.~\eqref{barh0_TT}, 
making use of the energy-momentum conservation $\partial_{\mu} T^{\mu \nu} = 0$ 
repeatedly to obtain the convenient relation 
\beq
2 \int d^3 x'\, T^{i j} 
= 
{\partial^2_{t}} \left( \int d^3 x'\, T^{00} {x'}^{i} {x'}^{j} \right)\,.
\eeq
Defining a (Newtonian) symmetric trace-free quadrupole moment by 
\beq
Q_{i j}(t) \equiv \int d^3 x' \rho (t, {\bf x}') 
\left( 
{x'}_{i} {x'}_{j} - \frac{1}{3} \delta_{i j}\, {{\bf x}'}^2 
\right)\,,
\label{defQ}
\eeq
where $\rho \equiv T^{00}$ is the Newtonian mass density 
and $\delta_{i j} \equiv {\rm diag} (1,\,1,\,1)$ is the Kronecker-delta. 
Inserting Eq.~\eqref{defQ} back into Eq.~\eqref{barh0_TT}, we arrive at 
\beq
{h}_{i j}^{\rm TT}(t,{\bf x}) = \frac{2}{r}\, \Lambda_{ij,\,kl}({\bf n})\, 
\frac{d^2 Q_{k l}}{d t^2} (t - r)\,.
\label{quadrupole_h}
\eeq
This is the well-known ``quadrupole formula'' for the GW signal. 

The two radiative degrees of freedom contained in ${h}_{i j}^{\rm TT}$  
can be conveniently extracted by introducing two unit polarization vectors 
${\bf p}$ and ${\bf q}$, which are orthogonal to the propagation direction ${\bf n}$ 
and to each other (satisfying $n_i n_j + p_i p_j + q_i q_j = \delta_{ij}$). 
The tensor ${h}_{i j}^{\rm TT}$ is then decomposed into 
two independent ``plus'' and ``cross'' polarization modes of GWs: 
\beq
{h}_{i j}^{\rm TT} = 
h_{+} (p_i p_j - q_i q_j) 
+ 
h_{\times} (p_i q_j + q_i p_j)\,.
\label{def-hpx}
\eeq
For example, if we use a Cartesian coordinate system ${\bf x} = (x,\,y,\,z)$, 
$h_{+}$ and $h_{\times}$ modes of GWs that propagate in the $z$-direction 
(so that ${\bf n} = (0,\,0,\,1)$) are 
\bea
\begin{array}{rcl}
h_+ (t, z) &=&\displaystyle \frac{1}{r} 
\left(
\frac{d^2 Q_{x x}}{d t^2} - \frac{d^2 Q_{y y}}{d t^2}
\right)\,,\\
h_{\times} (t, z) 
&=&\displaystyle \frac{2}{r} \frac{d^2 Q_{x y}}{d t^2}\,.
\label{hpx-z}
\end{array}
\eea

We now apply the GW polarization modes~\eqref{hpx-z} 
to the Newtonian binary system in a fixed circular orbit 
with the separation $R$ and the orbital frequency $\Omega$ 
(we will momentarily delay to discuss the back-reaction on the motion 
due to the GW emission from the system).
To compute the quadrupole moment $Q_{i j}$, 
we continue to use the Cartesian coordinate system 
and assume that the orbit lies in the $x$--$y$ plane 
whose center of mass is at the coordinate origin. 
In this setup, we have 
\beq
Q_{i j} = \mu \left( x_i x_j - \frac{R^2}{3} \delta_{i j} \right)
\label{circQ}
\eeq
with $x = R \cos (\Omega t + \pi / 2)$, $y = R \sin (\Omega t + \pi / 2)$ and $z = 0$ 
(so that the binary initially at $x(0) = 0$ and $y(0) = R$ when $t = 0$). 
The second derivative of Eq.~\eqref{circQ} is 
\bea
\frac{d^2 Q_{x x}}{d t^2} 
&=& 
2 \,\mu \,R^2 \Omega^2 \cos (2 \Omega\, t) 
= 
-\frac{d^2 Q_{y y}}{d t^2} 
\,, \cr
\frac{d^2 Q_{x y}}{d t^2} 
&=&
2 \,\mu \,R^2 \Omega^2 \sin (2 \Omega\, t)\,.
\label{ddotQ_circ}
\eea
Making substitution in Eq.~\eqref{hpx-z}, we obtain 
\bea
h_+(t) &=& 
\frac{4 \mu}{r} (\Omega R)^2 \cos (2 \Omega t_{\rm ret})
\,,
\cr
h_\times (t) &=& 
\frac{4 \mu}{r} (\Omega R)^2 \sin (2 \Omega t_{\rm ret})
\,,
\label{hpx0_fixed_circ}
\eea
where the retarded time $t_{\rm ret} \equiv t-r$ has been introduced.
We notice that the quadrupole radiation is at twice the orbital frequency $\Omega$.

From the observational point of view, the GW detector only sees the radiation 
in the direction that points from the binary to the detector, 
and it is not always the same as $z$-direction like Eq.~\eqref{hpx0_fixed_circ} in general.
When the GW propagates in a direction of the line of sight to the binary  
$
n_i \equiv (\sin \theta\, \sin \phi,\,\sin \theta \cos \phi,\,\cos \theta)
$, 
the GW polarization that an observer measures is conveniently rewritten in the form 
(see, e.g., Chapter~3 of Maggiore~\cite{Maggiore:1900zz})
\bea
h_+(t) &=& 
\frac{4 }{r}{\cal M}^{5/3} \,{\Omega}^{2/3} 
\left( \frac{1 + \cos^2 \theta}{2} \right) \cos (2 \Omega \,t_{\rm ret} + 2\phi)
\,,
\cr
h_\times (t) &=& 
\frac{4}{r}  {\cal M}^{5/3} \,{\Omega}^{2/3} 
\cos \theta \sin (2 \Omega \,t_{\rm ret} + 2\phi)
\,,
\label{hpx_fixed_chirp}
\eea
%where we introduce the retarded time $t_{\rm ret} \equiv t - r$, 
%and  the GW frequencies $2 \pi f_{\rm GW} \equiv 2 \Omega$, making 
where we use the Kepler's law for the circular orbit frequency $\Omega^2 = M / R^3$. 
Here, the combination of body's mass, the chirp mass:${\cal M}$, is defined by 
\beq
{\cal M} \equiv \eta^{3/5} M \,.
\label{def-chirpM}
\eeq
We remark that the amplitudes of the GW polarization at fixed $\Omega$
depend on the binary masses only through the chirp mass ${\cal M}$
(within the quadrupole approximation). 
%This parameter is well determined in the GW data analysis

%%%%%%%%%%%%%%%%%%%%%%%%%%%%%%%%%%%%%%%%%%%%%
\subsection{Adiabatic approximation}
%%%%%%%%%%%%%%%%%%%%%%%%%%%%%%%%%%%%%%%%%%%%%%

Our discussion has so far assumed that the GW is emitted from a given, fixed, 
(Newtonian) circular orbit. 
However, GW can transport energy, momentum, and angular momentum 
away from the binary, and hence the GW radiation actually drives an evolution 
of the circular orbit at the same time; 
the orbital separation $R$ will slowly shrink, causing the gradual radiative inspiral of the binary. 
In this subsection, we improve the GW signals~\eqref{hpx_fixed_chirp}, 
accounting for the inspiral motion of the binary due to the radiative losses.

The  total radiated power in all directions around the binary (i.e., the total energy flux of GWs)
that results from the quadrupole GW signal~\eqref{quadrupole_h} is in general given by 
(see, e.g., Chapter~12 of Poisson and Will~\cite{2014grav.book.....P} for a derivation)
\beq
F_{\infty}^{\rm Newt}
= 
\frac{1}{5} \sum_{i, j = 1}^{3} 
\left\langle\frac{d^3 Q_{i j}}{d t^3}\,\frac{d^3 Q_{i j}}{d t^3}\right\rangle\,,
\label{fluxN}
\eeq
where $\langle\dots\rangle$ stand for time average,
which is well-known as the quadrupole (``Newtonian'') formula for the GW energy flux. 
Inserting here the quadrupole moment of the circular binaries~\eqref{circQ}, 
we obtain 
\beq
F_{\infty}^{\rm Newt}
= 
\frac{32}{5} \eta^2 v^{10}\,,
\label{fluxN_circ}
\eeq
where $v \equiv \sqrt{M/R}=(M\Omega)^{1/3}$ is the (relative) orbital velocity. 
%and we assume $v \\ 1$ for large orbital separation

We now assume the so-called adiabatic approximation, in which a circular orbit 
evolves with a slowly changing orbital velocity $v$, so that its fractional change 
over an orbital period is negligibly small. 
In this approximation, the source of the GW energy fluxes 
comes from the sum of the kinetic and the (Newtonian) binding energy of the binary 
\beq
E^{\rm Newt} = -\frac{M \eta}{2} v^2 \,,
\label{eq:E_N}
\eeq
and hence it obeys the balance equation 
\beq
\frac{dE^{\rm Newt}}{dt} = - F_{\infty}^{\rm Newt} \,. 
\label{balanceN}
\eeq
The balance equation implies that the typical timescale of the radiation
reaction is 
\beq
t_{\rm RR} \equiv \frac{E^{\rm Newt}}{dE^{\rm Newt}/dt} \sim M\eta^{-1} v^{-8}\,,
\label{t_RR}
\eeq
which is much longer than the orbital time scale 
\beq
t_{\rm Orb} \sim M\, v^{-3}\,, 
\label{t_Orb}
\eeq
as far as the orbital velocity is small (or, the orbital separation is large), 
$v  = \sqrt{{M}/{R}} \ll 1$.
Therefore, the adiabatic approximation is mostly valid 
during the inspiral phase of the orbital evolution, 
but it breaks down as the binary separation approaches to the last stable orbit 
(whose typical orbital radius is close to that of the Innermost Stable Circular Orbit (ISCO) 
in Schwarzschild geometry with the mass $M$, $r_{\rm ISCO} = 6\, M$), 
where the transition from the inspiral phase to the merger phase approximately occurs.

In the adiabatic approximation, we can use the balance equation~\eqref{balanceN} 
to derive the evolution equation for any binary parameters. 
For instance, we have the evolution equation of the orbital velocity $v(t)$,
\beq
\frac{dv}{dt} = 
\frac{32}{5} \frac {\eta}{M}\, v^9\,.
\label{eq:dvdt_L}
\eeq
The time to coalescence $t_c$ is formally defined by the time it takes the velocity to evolve from an initial value $v_{\rm Ini}$ to $v \to \infty$, and it is
\beq
t_c 
= \frac {5}{256} {\frac {M}{\eta\,v_{\rm Ini}^{8}}} 
=\frac {5}{256}
\left(\frac{R_{\rm Ini}}{M}\right)^4 \frac{(1+q)^2}{q} M \,,
\label{T-of-coalescence}
\eeq
where $R_{\rm Ini} = M v^{-2}_{\rm Ini}$ denotes the initial orbital velocity and radius, respectively.
Similarly, the differential equation
\beq
\frac{dR}{dt} 
= \left(\frac{dE}{dR} \right)^{-1} \frac{dE}{dt} 
= -\frac{64}{5} \frac{\eta M^2}{R^2} \,,
\eeq
determines the shrinking rate of the orbital separation $R(t)$.
Assuming $r(t_c)=0$, we have the solution 
\beq
R(t) = \left( \frac{256}{5} \, \eta\, M^3\right)^{1/4} (t_c - t)^{1/4} \,.
\eeq
In turn, recall the Kepler's third law $\Omega^2 = M /R^3$, 
it leads the increasing orbital frequency  
\beq
\Omega(t) 
=
\left(\frac{5}{256} \frac{1}{t_c - t} \right)^{3/8} \, {\cal M}^{-5/8}\,.
\label{N_Omega_t}
\eeq
Importantly, the frequency evolution depends on the binary masses only through 
the chirp mass ${\cal M}$ in the zero-th order approximation
leading to Eq.~(\ref{quadrupole_h}).

We next wish to describe the impact of these time-dependent binary parameters 
on the GW signals~\eqref{hpx_fixed_chirp} within the adiabatic approximation. 
Recall the discussion of the quadrupole GW generation formalism in the preceding subsection, 
the orbital frequency $\Omega$ and the GW phase $2\Omega\, t_{\rm ret}$ have to be promoted to
\beq
\left\{ 
\Omega ,\, 2 \Omega \, t_{\rm ret}
\right\} 
\to 
\left\{ 
\Omega(t_{\rm ret}) ,\, \phi(t_{\rm ret})
\right\}\,.
\label{fix-to-insp}
\eeq
Here, we introduce a new accumulated phase of GW 
(associated with the time-dependent orbital frequency~\eqref{N_Omega_t}) by 
\beq
\phi(t) \equiv 2 \int \Omega(t) dt
= - 2 \left(\frac{1}{5} \frac{t_c - t}{\cal M} \right)^{5/8}
+ \phi_c \,,
\label{inspi_phi}
\eeq
and $\phi_c \equiv \phi(t = t_c)$ is the phase at the time to coalescence $t_c$. 
Inserting Eq.~\eqref{fix-to-insp} 
%with Eqs.\eqref{N_Omega_t} and~\eqref{inspi_phi}
into Eq.~\eqref{hpx_fixed_chirp}, we arrive at 
\bea
h_+(t) &=& 
\frac{ {\cal M}}{r} \left( \frac{5 {\cal M}}{t_c - t} \right)^{1/4} 
\left( \frac{1 + \cos^2 \theta}{2} \right) \, 
\cos \left\{- 2 \left(\frac{1}{5} \frac{t_c-t}{\cal M} \right)^{5/8}
+ \phi_c \right\}
\,,
\cr
h_\times (t) &=& 
\frac{ {\cal M}}{r}  \left( \frac{5 {\cal M}}{t_c - t} \right)^{1/4} 
\cos \theta \, \sin \left\{- 2 \left(\frac{1}{5} \frac{t_c-t}{\cal M} \right)^{5/8}
+ \phi_c \right\}
\,.
\label{hpx_chirp}
\eea
This is the GW polarization from inspiring quasicircular binaries 
that an observer measures in a quadrupole approximation. 
Both the amplitudes and the frequency in Eq.~\eqref{hpx_chirp} increase
as the coalescence time is approached: this is referred to as ``chirping'' 
(in fact, the ``chirp mass'' $\cal M$ is named after it), 
and we often call Eq.~\eqref{hpx_chirp} the chirp signal.
Also, we note that both the amplitude and the phase still depend on the binary masses 
only through the ${\cal M}$. 
This explains why the chirp mass is well determined in the GW data analysis, 
compared with the component masses of the binary, $m_a$.

%%%%%%%%%%%%%%%%%%%%%%%%%%%%%%%%%%%%%%%%%%%%%%%%%%
\subsection{Stationary phase approximation}
%%%%%%%%%%%%%%%%%%%%%%%%%%%%%%%%%%%%%%%%%%%%%%%%%%

An alternative to the time-domain chirp signal~\eqref{hpx_chirp} 
is its frequency-domain (i.e., Fourier domain) representation, 
which is also commonly used for the GW data analysis applications.
The time-domain chirp signal takes the schematic 
complex-exponential form of $h(t) = A(t) e^{- i \phi(t)}$, 
and its Fourier transform is 
\beq
{\tilde h} (f) = \int d t \,A(t) \, e^{i (2 \pi f t - \phi(t))} \,;
\label{def-Fourier}
\eeq
note that ${\tilde h}(-f) = {\tilde h}^{\ast} (f)$, so we can assume $f > 0$. 
Since the amplitude of the chirp signal evolves much slower than the phase, i.e., 
\beq
\frac{d \ln A(t)}{dt} \ll \frac{d \phi(t)}{dt} \,,
\eeq
the stationary phase approximation to the integral provides 
a good approximation of ${\tilde h} (f)$:
\beq
{\tilde h} (f) \simeq \frac{A (t)}{\sqrt{dF/dt (t_*)}}\, 
e^{i \, (\Psi_{\rm SPA}(t_*) - \pi / 4)}\,, 
\label{def-SPA}
\eeq
where 
\beq
\Psi_{\rm SPA} (t_*) \equiv 2 \pi f t_* - \phi(t_*)\,,
\eeq
and $t_*$ is a function of $f$ defined as
\beq
\frac{d \phi(t_*)}{dt_*} \equiv 2 \pi f \,,
\eeq
and the time at the stationary point $t_*$ is determined by the Fourier
variable $f$ being equal to the instantaneous frequency $d\phi(t)/dt$ at $t=t_*$,

The explicit calculation of the stationary phase approximation 
to the time-domain chirp signals~\eqref{hpx_chirp} is worked out 
in, e.g., Chapter~4 of Maggiore~\cite{Maggiore:1900zz}. 
The resultant frequency-domain chirp signal is  
\bea
{\tilde h}_+(f) &=& 
\frac{1}{\pi^{2/3}} \sqrt \frac{5}{24} \frac{{\cal M}^{5/6}}{f^{7/6}}
\frac{e^{i \, \Psi_+(f)}}{r}\, 
\left( \frac{1 + \cos^2 \theta}{2} \right) \,,
\cr
{\tilde h}_\times (f) &=& 
\frac{1}{\pi^{2/3}} \sqrt \frac{5}{24} \frac{{\cal M}^{5/6}}{f^{7/6}}
\frac{e^{i\, \Psi_{\times}(f)}}{r}\, \cos \theta \,,
\label{SPA_chirp}
\eea
where the phases are given by
\bea
\Psi_{+}(f) &&
\equiv
2 \pi f (t_c) - \phi_c - \frac{\pi}{4} + \frac{3}{128} (\pi {\cal M} f )^{-5/3}\,,
\cr
\Psi_{\times}(f)  &&\equiv \Psi_{+}(f) + \frac{\pi}{2}\,.
\label{SPA_phase}
\eea

%%%%%%%%%%%%%%%%%%%%%%%%%%%%%%%%%%%%%%%%%%%%%%%%%%%%%%%%%%%%%%%%%%%%%%%%%%%%%%%%%%%%%%%%%%%%%%%%%%%
\section{Post-Newtonian gravitational waveforms for spinning, nonprecessing binary black holes }
%%%%%%%%%%%%%%%%%%%%%%%%%%%%%%%%%%%%%%%%%%%%%%%%%%%%%%%%%%%%%%%%%%%%%%%%%%%%%%%%%%%%%%%%%%%%%%%%%

In our discussion so far, we have restricted our analysis 
to the Newtonian orbital dynamics in the linearized gravity theory 
(with the quadrupole formula for the GW fluxes). 
While this treatment has provided an adequate description 
of the dynamics of the binary pulsars (e.g., the secular change in the orbital period 
of the Hulse-Taylor binary: PSR B1913$+$16; see a review~\cite{Lorimer:2005bw}), 
it is not accurate modelling of the GW signals emitted from inspiralling 
astrophysical binary system; 
the chirp signals~\eqref{hpx_chirp} are just inconsistent with the observed GW data of, e.g., 
BBHs GW150914~\cite{Abbott:2016blz} and 
BNS GW170817~\cite{TheLIGOScientific:2017qsa}.    
For the latter case, one must rely on an improved approximation method 
in GR to have a more refined wave-generation formalism.

The perfect starting point for that discussion is the ``post-Newtonian (PN) theory,'' 
applied to the two-body problem. 
The PN theory is a systematic approximation method to exact GR, 
solving the Einstein field equation (and the equation of the motion for a source) 
in the form of power series in small physical parameters 
\bea
v \ll 1 \,,
\quad
\frac{M}{R} \sim {v^2} \ll 1 \,,
\label{PNparam}
\eea
namely, the two-body system is assumed to move slowly (with a large separation), 
and to be in the weak gravitational field. 
Therefore, the PN theory provides the most natural tool to model 
the early inspiral stage for LIGO-Virgo-KAGRA binary mergers.  

The technical developments of the general wave-generation formalism 
in the PN theory is far more involved than the quadrupole moment formalism  
in the linearized theory, and we shall continue to specialise our discussion  
to GW signals emitted by a binary system in the (slowly evolving) circular orbit 
for simplicity.
We refer the reader the text by Poisson and Will~\cite{2014grav.book.....P}, 
Blanchet's Living review article~\cite{Blanchet:2013haa},
and Chapter~32 by Sturani in this book for the effective field theory approach.

In the PN theory, the GW polarizations $h_{+ ,\, \times}$ produced 
by a circular binary system have the following general structure 
($p$ is an integer number)
%~\cite{Blanchet:2013haa}: 
\beq
h_{+,\,\times} 
=
\frac{2 \, \mu \, v^2}{r}\, \sum_{p\geq 0} v^{p}\, H^{(p)}_{+,\,\times}
+ O\left( \frac{1}{r^2} \right)\,,
\label{PN-hpx}
\eeq
where the variable $v$ that was a relative orbital velocity in the previous section 
is conveniently redefined as the frequency-related parameter by  
\beq
v^2 \equiv (M \Omega)^{2/3} =
\frac{M}{R} \left\{ 1 + O\left( \frac{1}{c^2} \right) \right\}
\,. 
\label{def-vPN}
\eeq
The leading-order terms of Eq.~\eqref{PN-hpx} explicitly reads 
(ignore the static non-linear memory contribution 
to $H^{(0)}_{+}$~: see, e.g., Chapter~9 of Ref.~\cite{Blanchet:2013haa}) 
\beq
H^{(0)}_{+}
=
- (1 + \cos^2 \theta) \cos 2 \psi\,, 
\quad
H^{(0)}_{\times}
=
-2 \cos \theta \sin 2 \psi\,.
\label{H0}
\eeq
Here, we introduce the ``tail-distorted'' phase, 
\beq
\psi = \phi(t) - 6 v^3 \ln v \,,
\label{def-psi}
\eeq
where the binary's orbital phase $\phi(t)$ receives a
correction from the scattering of the GW off the static curvature generated by
the binary itself, note that since $\phi(t)\sim v^{-5}$ in the quadrupole approximation, see Eqs.~\eqref{N_Omega_t} and \eqref{inspi_phi}, this is a relative
$v^8$ correction to the leading order.

It is convenient to decompose the GW polarizations~\eqref{PN-hpx} 
onto the (spin-weighted) spherical harmonic mode 
(see, e.g., Section~3.1 of Blanchet's review~\cite{Blanchet:2013haa})  
when comparing the PN waveform with, e.g., the numerical-relativity waveform. 
This is often called GW modes $h_{\ell m}$, which is expressed as 
\beq
h_+(t) - i\, h_\times(t)
= \sum_{\ell=2}^{\infty}\, \sum_{m=-\ell}^{\ell}
h_{\ell m}(t)\,{}_{-2} Y_{\ell m}(\theta,\phi) \,,
\label{hlm}
\eeq
where ${}_{-2} Y_{\ell m}$ is the spin ($-2$)-weighted spherical harmonics, 
and we note that $h_{\ell-m}=(-1)^\ell {\bar h}_{\ell m}$. 
The dominant quadrupole GW modes $(\ell,\,m) = (2,\,2)$ is (see, e.g., Ref.~\cite{Ajith:2012az}) 
\beq
h_{22} = \frac{8 \mu}{r} \sqrt{\frac{\pi}{5}} v^2 e^{- i 2 \psi}
+ O\left( v^3,\,\frac{1}{r^2} \right)\,,
\label{h22}
\eeq
and the other GW modes starts at $O(v^3)$ or higher.

For the GW data analysis applications, the phasing of the GW signal is significantly more important 
than its amplitude due to the matched-filter searching of the GW signal.  
Thus, the (so named) ``restricted'' PN waveform~\cite{Cutler:1992tc} is commonly used,  
in which only leading term $H^{(0)}_{+,\,\times}$ (or $h_{22}$) of the waveform is retained,  
while all the PN correction to the orbital phase evolution $\phi(t)$ in Eq.~\eqref{def-psi} 
are included.
(we note, however, that the higher-order amplitude terms become more pronounced especially 
for precessing or unequal mass-ratio binaries, e.g., GW190412~\cite{LIGOScientific:2020stg}). 
Because the early inspiral phase of binary is in the adiabatic regime, 
where the typical radiation-reaction time scale is much longer than the typical orbital time scale 
$t_{\rm Orb}/t_{\rm RR} \sim O(v^5) \ll 1$ 
(recall Eqs.~\eqref{t_RR} and~\eqref{t_Orb}),  
the orbital phase evolution $\phi(t)$ in the restricted PN approximation 
can be computed efficiently, making use of the adiabatic approximation discussed previously.

The orbital phase $\phi(t)$ nonetheless has to be calculated up to a very-high-order PN term. 
The PN phase evolution can be parametrized by the following general structure
(discard the constant phase): 
\beq
\phi(v) 
=
-\frac{1}{32 \, \eta} \frac{1}{v^5}
\left\{ 1 + O(v^2) + O(v^3) + O(v^4) + O(v^5) + \dots \right\} \,.
\eeq
Because the leading term in $\phi(v)$ scales like $O(v^{-5})$, 
one needs to compute the PN corrections at least $2.5$PN order (or even higher order) to keep the absolute phase error to less than $O(1)$,
which is needed for the effectiveness of the matched-filtering search.

Throughout the rest of this section, assuming the adiabatic approximation 
and making a further specialization to the BBHs with aligned spins 
and without orbital eccentricity, we will review how the orbital phase $\phi(t)$ can be computed 
up to the $3.5$PN order~\cite{Isoyama:2017tbp};
adding the neglected effects for other binary configurations will be discussed in the next section. 
In this simpler setup, we will just need i) the PN (center-of-mass) binding energy, 
ii) the energy flux emitted to the infinity and across the BH horizons, 
and iii) the certain balance equation associated with them, as inputs.

%%%%%%%%%%%%%%%%%%%%%%%%%%%%%%%%%%%%%%%%%%%%%%%%%%%%%%%%%%%%%%%%%%%%
\subsection{PN binding energy, energy flux, and BH horizon flux}
%%%%%%%%%%%%%%%%%%%%%%%%%%%%%%%%%%%%%%%%%%%%%%%%%%%%%%%%%%%%%%%%%%%%

We schematically express the $3.5$PN corrections 
to the Newtonian (center-of-mass) binding energy~\eqref{eq:E_N} as 
\beq\label{def-E}
E \equiv 
-\frac{M \eta}{2} v^2 
\left\{
E_{\rm {NS}}
+
\frac{v^3}{M^2} E_{\rm {SO}}
+
\frac{v^4}{M^4} E_{\rm {SS}} 
+
\frac{v^7}{M^6} E_{\rm {SSS}} 
+ 
O(v^8)
\right\}\,,
\eeq
where ``NS,'' ``SO,'' ``SS,'' and ``SSS''
denote the non-spinning, 
spin-orbit (linear-in-spin),
spin-spin (quadratic-in-spin),
and spin-spin-spin (cubic-in-spin) contributions,
respectively; 
the explicit expressions for $E_{\rm {NS}}$ are given in, e.g., Ref.~\cite{Blanchet:2013haa} 
while those for $E_{\rm {SO}},\,E_{\rm {SS}}$ and $E_{\rm {SSS}}$ are given 
in, e.g., Ref.~\cite{Bohe:2012mr} and~\cite{Marsat:2014xea}, 
respectively (see also Refs.~\cite{Mishra:2016whh} and references therein).
In the parenthesis of the right-hand side of the above equation, we have factored out $v$ and $M$
where $v$ shows the leading PN order of each term
and $M$ is related to the powers of the spins.
We note that the complete binding energy is also available up to 4PN order 
both in the spinning sector~\cite{Levi:2016ofk} and 
the non-spinning sector~\cite{Damour:2015isa,Jaranowski:2015lha,Bernard:2017ktp,Foffa:2019rdf}, 
the latter of which has very recently been obtained: 
\bea
E^{\rm 4PN}_{\rm NS}
&=&
v^8
\left[
-\frac{3969}{128} + 
\left\{
-\frac{123671}{5760} + \frac{9037}{1536}\pi^2 + \frac{896}{15} \gamma_{\rm E} 
+\frac{448}{15} \ln (16 v)
\right\}\,\eta
\right. \cr
&& \left.  \quad
+
\left(
-\frac{498449}{3456} + \frac{3157}{576} \pi^2
\right)\, \eta^2
+
\frac{301}{1728} \eta^3 + \frac{77}{31104} \eta^4
\right]\,.
\label{ENS_4PN}
\eea
%The significant progress has been made in pushing the state-of-the-art 
%with the current frontier at the 4.5PN and the 5PN: 
%see, e.g.,Refs.~\cite{Foffa:2019eeb} and reference therein.

Similarly, the $3.5$PN corrections to the Newtonian energy flux~\eqref{fluxN_circ}
emitted to the infinity is written as
\beq\label{def-Finf}
{F}_{\infty} \equiv 
\frac{32}{5} \eta^2 v^{10} 
\left\{
{F}_{\mathrm {NS}}
+
\frac{v^3}{M^2} {F}_{\mathrm {SO}}
+
\frac{v^4}{M^4} {F}_{\mathrm {SS}} 
+
\frac{v^7}{M^6} {F}_{\mathrm {SSS}}
+
O(v^8)
\right\}\,.
\eeq
The explicit expressions for $F_{\rm {NS}}$ and ${F}_{\mathrm {SO}}$
are given in, e.g., Ref.~\cite{Blanchet:2013haa} while that for 
$F_{\rm {SS}}$ and $F_{\rm {SSS}}$ given in, e.g., Ref.~\cite{Bohe:2012mr} 
and~\cite{Marsat:2014xea}, respectively (see also Refs.~\cite{Mishra:2016whh}
and references therein).
Currently, the complete form of GW fluxes beyond $3.5$PN order is missing  
(only relative $O(\eta)$ piece is available up to $11$PN order~\cite{Fujita:2014eta}), 
and its derivation is a frontier in the PN calculations.

When the coalescing binary has at least one BH component, 
there is a part of the GW flux that goes down to the BH horizon,  
due to absorption effects of the energy and angular-momentum GW fluxes
across the BH horizon. 
Such horizon flux appear at $2.5$PN order for spinning BHs~\cite{Tagoshi:1997jy}
(but it is pushed to $4$PN order for non-spinning BHs~\cite{Poisson:1994yf,Goldberger:2020wbx})
relative to the leading Newtonian-order energy flux~\eqref{fluxN_circ}, 
and it is known up to relative $1.5$PN order for arbitrary mass ratio~\cite{Chatziioannou:2016kem}
(i.e., 4PN order beyond leading order energy flux)
as well as $11$PN order for the linear in the mass-ratio part~\cite{Fujita:2014eta}, 
beyond the leading-order fluxes. 
The horizon fluxes (absorbed by the BH labelled by $a$) may have 
the following factorized from  
\beq
{F}_{\rm H}^{a}(t\,;m_a,\chi_a) 
=
\Omega_{\rm tidal} (\Omega_{H} - \Omega_{\rm tidal})\, C_v^a\,,
\label{def-FH}
\eeq
where
\beq
C_v^a
\equiv
-\frac{16}{5}\, \frac{m_a^4}{M^2} \eta^2 
\left(1 + \sqrt{1 - \chi_a^2}\right)\,v^{12}
\left\{
1 + O(v^2) + \dots
\right\}\,.
\eeq
Here, $\Omega_{\rm tidal} = O(v^3)$ and $\Omega_{\rm H}$ are the angular velocities of 
the tidal field (caused by a companion BH) and the BH horizon, respectively. 
The coefficient $C_v^a$ denotes the remaining factor determining the horizon flux, 
and it is needed to next-to-leading (relative $1$PN) order to achieve $3.5$PN order
precision.

%%%%%%%%%%%%%%%%%%%%%%%%%%%%%%%%%%%%%%%%%%%%%%%%%%%%%%%%%%%%%%%
\subsection{Balance equation for slowly evolving black holes}
%%%%%%%%%%%%%%%%%%%%%%%%%%%%%%%%%%%%%%%%%%%%%%%%%%%%%%%%%%%%%%%

The most important equation in the adiabatic approximation, 
analog to Eq.~\eqref{balanceN} in linearized theory, 
is the energy balance equation. 
In the case of BBHs, it is written as 
\beq\label{balance0}
\frac{dE}{dt} = - F_{\infty} - \sum_a F_{\rm H}^a\,,
\eeq
where the change rate of the center-of-mass binding energy $E$ 
(related to conservative dynamics), 
is equated to the energy fluxes of the GW emission carried out 
to infinity $F_{\infty}$ and down to the BH horizon $F_{\rm H}^a$ 
(related to dissipative dynamics).
A key assumption made in Eq.~\eqref{balance0} is that the BH masses $m_a$ and 
BH spins $S_a$ remain constant in the inspiral phase 
until the adiabatic approximation itself breaks down. 

However, this assumption is violated by horizon absorption effects. 
Each BH can be slowly evolving at the same time by changing its mass and spin 
via horizon absorption, namely 
\beq
\frac{d m_a}{dt} = F_{\rm H}^a\,,
\quad
\frac{d S_a}{dt} = \frac{1}{\Omega_{\rm tidal}}\,\frac{d m_a}{dt}\,.
\label{dmdt}
\eeq
The absorption-corrected energy balance equation up to 3.5PN order 
%that accounts for slowly evolving component BHs 
is schematically given by Ref.~\cite{Isoyama:2017tbp}
\bea\label{balanceH-0}
\left( \frac{\partial {\cal E}}{\partial t} \right)_{m,\,S} 
= 
- {\cal F}_{\rm eff}
\equiv
-{\cal F}_{\infty} - \sum_{a} ( 1 - \Gamma_{\mathrm H}^a )
\,{\cal F}_{\mathrm H}^a\,,
\eea
where the absorption-corrected binding energy ${\cal E}$ 
and fluxes ${\cal F}_{\infty,\, {\rm H}}$ are defined 
by corresponding $E$ and ${F}_{\infty,\, {\rm H}}$, 
promoting the constant BH mass $m_a$ and spin $S_a$ 
to the slowly evolving mass $m_a(t)$ and spin $S_a(t)$. 
The BH's growth factors $\Gamma_{\mathrm H}^a = O(v^2)$ account 
for the fact that the derivative on the left hand is now partial one.

%%%%%%%%%%%%%%%%%%%%%%%%%%%%%%%%%%%%%%%%%%%%%%%%%%%%%%%%%%%%
\subsection{Accuracy of the post-Newtonian approximants}

At this point, we must be mindful of the accuracy of the PN approximations.
Two possible approaches to this problem have been extensively investigated in the literature. 
The first approach is to directly compare the PN waveform against 
the exact waveform from full NR simulations, performed in, e.g., Ref.~\cite{Szilagyi:2015rwa}. 
The second approach is to compare physical quantities computed,  
such as the GW energy fluxes and binding energy (as well as linear momentum) 
that can be computed at higher PN order than the waveforms themselves: 
the modern development on both approaches is nicely summarised 
by Le Tiec's review~\cite{Tiec:2014lba}.

This subsection is an additional contribution to the latter, 
based on the recent comparison of the GW energy fluxes~\cite{Sago:2016xsp}. 
The general strategy is to rely 
on Black Hole Perturbation (BHP) theory 
(see, e.g., Refs. ~\cite{Nakamura:1987zz,Mino:1997bx,Sasaki:2003xr} 
for reviews as well as Chapter 36 by Pound and Barry in this book) 
outlined by earlier work in Refs.~\cite{Yunes:2008tw,Zhang:2011vha} 
in which very high PN order calculations can be achieved systematically
by further expanding the PN series in the mass ratio $q$ assumed to be small; 
see also ``Black Hole Perturbation Toolkit'' at \url{http://bhptoolkit.org/}
and ``Black Hole Perturbation Club'' at \url{https://sites.google.com/view/bhpc1996/home} 
for further details of the BH perturbation theory. 

Consider a small body moving along the quasi equatorial-circular orbit in the Kerr spacetime 
with the mass $M$ (note that $M$ is not the total mass
in this subsection) and Kerr spin parameter $a/M$.
Assuming a ``small-mass-ratio'' limit ($q \to  0$), the $11$PN GW energy flux
radiated to the infinity at the leading order in $q$
has been computed~\cite{Fujita:2014eta}:  
\beq
 F^{(N)} = \sum_{k=0}^{N} \sum_{p=0}^{[k/6]} F^{(k,p)} \left\{\ln(v)\right\}^p v^k \,,
\label{eq:F_BHPC}
\eeq
where $p$ and $k$ are integer numbers and $N=22$, 
and the leading order has been to normalized to unity: $F^{(0,0)}=1$.  

\begin{figure}[th]
    \centering
    \includegraphics[width=0.8\textwidth]{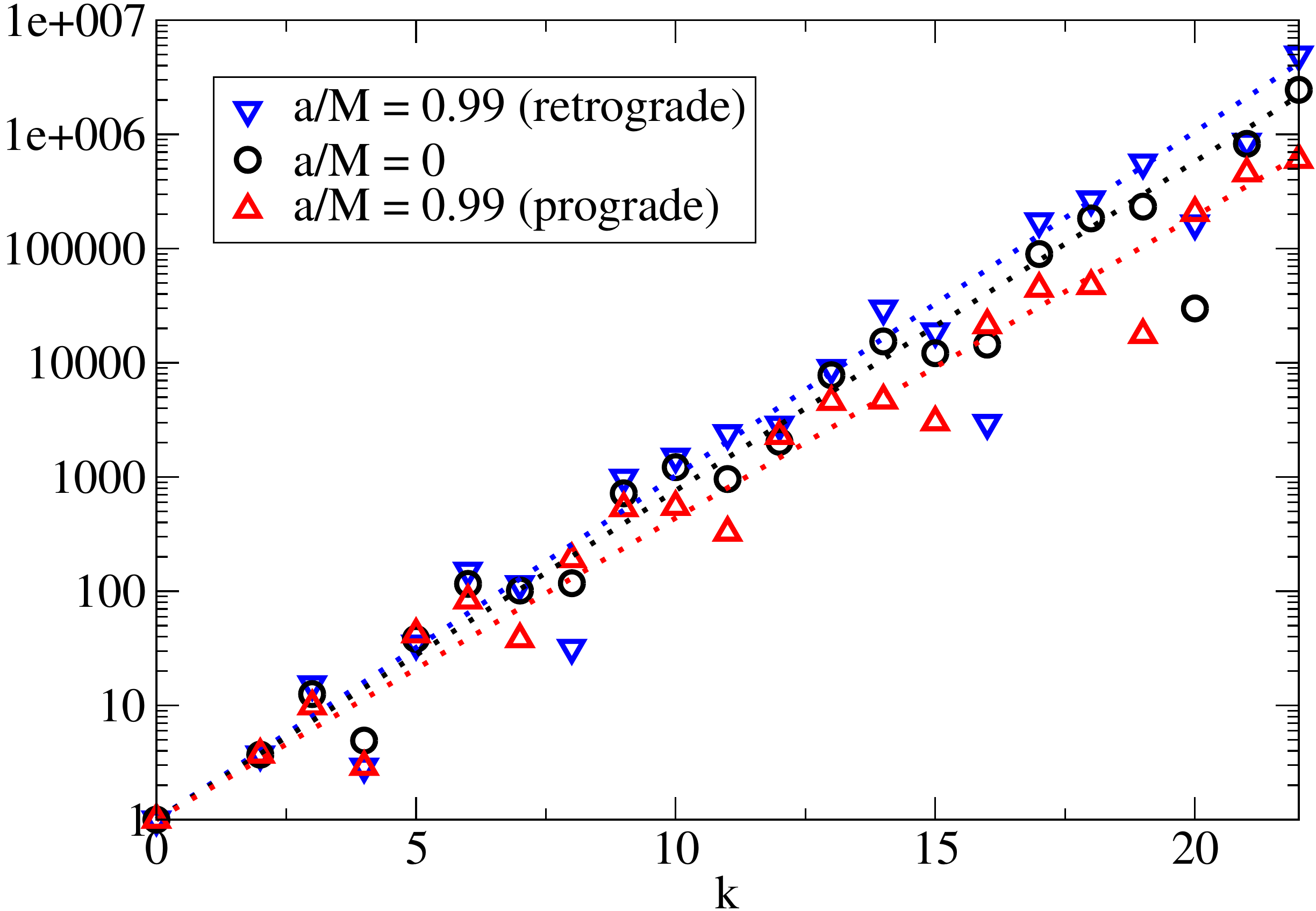}
    \caption{The inverted (blue) triangles, (black) circles,
    and (red) triangles denote $F^{(k,0)}$
    in Eq.~\eqref{eq:F_BHPC}
    for the Kerr spin parameter $a/M=0.99$ (retrograde orbit), 
    $0$, and $0.99$ (prograde orbit), respectively.
    The dashed lines are the fitting 
    and show $(2.00014)^k$ for $a/M=0.99$ (retrograde orbit),
    $(1.94120)^k$ for $a/M=0$, and $(1.83690)^k$ for 
    $a/M = 0.99$ (prograde orbit).}
    \label{fig:conv}
\end{figure}

The PN coefficients $F^{(k,p)}$ are approximately fitted
by a linear function with respect to $k$ in the log-linear plot. 
As a specific example, the PN coefficient $F^{(k,0)}$ versus $k$
in the cases of the Kerr spin parameters of $a/M=0.99$
(for the retrograde orbit), $0$, and
$0.99$ (for the prograde orbit) is displayed in Fig.~\ref{fig:conv}. 
The inverted (blue) triangles, (black) circles, 
and (red) triangles denote $F^{(k,0)}$ in the three cases, respectively.
The dashed lines are the fitting lines for each case,
and we find that $F^{(k,0)}$ ($k=0, \ldots, 22$) are fitted as $(2.00014)^k$ 
for $a/M =0.99$ (retrograde), $(1.94120)^k$ for $a / M = 0.0$, and $(1.83690)^k$ 
for $a/M = 0.99$ (prograde).
These fittings suggest that the approximate radius of convergence 
in terms of the velocity parameter $v$ is expected to be 
\bea
v_{\rm conv} (a/M =0.99) &\sim& 0.499965\, {\rm (retrograde)}\,; \cr
v_{\rm conv} (a/M =0.00) &\sim& 0.515145\,; \cr
v_{\rm conv} (a/M =0.99) &\sim& 0.544395\, {\rm (prograde)}\,.
\label{v_conv}
\eea

These values should be compared with the frequency at the innermost stable circular orbit (ISCO) 
because the quasicircular inspiral of a small body lasts until the orbital separation shrinks 
to the ISCO radius $r_{\mathrm {ISCO}}$ (in the Boyer-Lindquist coordinate).
The ISCO radius in the equatorial plane of the Kerr spacetime is given by~\cite{Bardeen:1972fi}
\beq
\label{eq:ISCO_radius}
r_{\mathrm {ISCO}} = 
M \left[ 3 + Z_2 \mp \left\{(3 - Z_1) (3 + Z_1 + 2 Z_2)\right\}^{1/2} \right] \,,
\eeq
where $Z_1 \equiv 1 + (1 - \chi^2)^{1/3} \{(1 + \chi)^{1/3} + (1 - \chi)^{1/3}\}$ 
and $Z_2\equiv (3 \chi^2 + Z_1^2)^{1/2}$.
Here, the upper/lower sign refers to prograde/retrograde orbits. 
In Table~\ref{tab:ISCO}, we summarize the ISCO radius ($r_{\rm ISCO}/M$),
ISCO frequency ($M\, \Omega_{\rm ISCO} = \{(r_{\rm ISCO} / M)^{3/2} + \chi\}^{-1}$) 
and the frequency parameter at ISCO ($v_{\rm ISCO} \equiv (M \Omega_{\rm ISCO})^{1/3}$)
for the representative Kerr spin parameters.
Comparing the values of $v_{\rm conv}$~\eqref{v_conv} and $v_{\rm ISCO}$ in Table~\ref{tab:ISCO},
we expect that the PN series to the GW energy fluxes would work well 
up to ISCO for most of the retrograde orbits, 
but not for the prograde orbit with the high BH spin. 
%---------------------------------------------------------------
\begin{table}[ht]
\caption{The ISCO radius ($r_{\rm ISCO}/M$), ISCO frequency ($M\,\Omega_{\rm ISCO}$) 
and the frequency parameter a ISCO ($v_{\rm ISCO}$)
for the representative Kerr spin parameters ($a/M$); 
compare them against the values of the approximate radius of convergence 
$v_{\rm conv}$ in Eq.~\eqref{v_conv}. 
%obtained from Fig.~\ref{fig:conv}
%is expected to be $v_{\rm conv}=0.499965$, $0.515145$, and $0.544395$ for $a/M=0.99$ (retrograde),
%$a/M=0$, and $a/M=0.99$ (prograde), respectively.
}
\label{tab:ISCO}
\begin{center}
\begin{tabular}{c|ccc}
$a/M$ & $r_{\rm ISCO}/M$ & $M\Omega_{\rm ISCO}$ & $v_{\rm ISCO}$ \\
\hline
1.0 (retrograde) & 9.000000000 & 0.03571428571 & 0.3293168780
\\
0.9 (retrograde) & 8.717352279 & 0.03754018063 & 0.3348359801
\\
0.5 (retrograde) & 7.554584713 & 0.04702732522 & 0.3609525320
\\
0.0 & 6.000000000 & 0.06804138173 & 0.4082482904
\\
0.5 (prograde) & 4.233002531 & 0.1085883589 & 0.4770835292
\\
0.9 (prograde) & 2.320883043 & 0.2254417086 & 0.6086179484
\\
1.0 (prograde) & 1.000000000 & 0.5000000000 & 0.7937005260
\\
\end{tabular}
\end{center}
\end{table}
%--------------------------------------------------------------- 

%%%%%%%%%%%%%%%%%%%%%%%%%%%%%%%%%%%%%%%%%%%%%%%%%%%%%%%%%
\section{Time- and Frequency-Domain Inspiral Templates}
%%%%%%%%%%%%%%%%%%%%%%%%%%%%%%%%%%%%%%%%%%%%%%%%%%%%%%%%%

In this section we shall construct PN GW templates 
for the early adiabatic inspiral of spinning, nonprecessing BBHs, 
including the secular evolution of BH mass and spin. 
Our main goal is to obtain the phase function $\phi(t)$~\eqref{def-psi} to $3.5$PN order, 
making use of the (BH-absorption corrected) PN binding energy~\eqref{def-E}, 
PN energy fluxes~\eqref{def-Finf} and~\eqref{def-FH}, 
and the generalized balance laws for the slowly evolving BHs~\eqref{balanceH-0} 
introduced in the previous section. 
The obtained phase in this way are called Taylor 
PN approximants~\cite{Damour:2000zb,Damour:2002kr,Buonanno:2009zt}. 
For time-domain templates, we will present TaylorT1, TaylorT2, TaylorT3, TaylorT4 
and TaylorT5 approximants.
For the frequency-domain templates, we will show TaylorF1 and TeylorF2 approximants.
These Taylor approximants are formally equivalent up to the $3.5$PN order, 
but the uncontrolled (higher-order) PN order terms are truncated differently. 

Throughout the rest of this section, the labels 
``NS'', ``SO'', ``SS'' and ``SSS''
denote the spinning, point-particle's contributions, namely, without BH absorption of
non-spinning, spin-orbit (linear-in-spin),
spin-spin (quadratic-in-spin),
and spin-spin-spin (cubic-in-spin) terms to the phase, 
while all the BH-absorption corrections are labelled 
by ``Flux, 5'' (from LO ($2.5$PN) horizon flux), 
``Flux, 7'' (from NLO ($3.5$PN) horizon flux), 
and ``BH, 7'' (from slowly evolving, BH mass and spin).
The explicit expressions for the point-particle contributions, not including
absorption effect, are 
implemented in LALSimulation~\cite{lalsuite} (see also Ref.~\cite{Buonanno:2009zt})
as ``Module LALSimInspiralTaylorXX.c''at \url{https://lscsoft.docs.ligo.org/lalsuite/lalsimulation/group___l_a_l_sim_inspiral_taylor_x_x__c.html},
while all the BH-absorption contributions are listed in Ref.~\cite{Isoyama:2017tbp}.

Before proceeding, we note that our construction in this section is not complete. 
The frequency evolution due to changes in the BH's mass and spin 
are constrained by (adiabatically invariant) ``first-laws'' 
of compact binary mechanics~\cite{LeTiec:2011ab,Blanchet:2012at,Fujita:2016igj}, 
which will be disregarded here; 
accounting for the first-law effect will await future work.
See the recent work by Hughes~\cite{Hughes:2018qxz} for details.

%%%%%%%%%%%%%%%%%%%%%%%%%%%%%%%%%%%%%%%%%%%%%%%%%%%
\subsection{Taylor time domain approximants}
%%%%%%%%%%%%%%%%%%%%%%%%%%%%%%%%%%%%%%%%%%%%%%%%%%%

Using the balance equations~\eqref{balanceH-0},  
the evolution equations of the phase $\phi(t)$ are~\cite{Isoyama:2017tbp}
\bea
\label{Kepler}
\frac{d \phi}{d t} &=& \Omega = \frac{v^3}{M(v)} \,, \\ 
\label{dvdt}
\frac{d v}{d t} 
&=&
-
\frac{{\cal F}_{\mathrm {eff}}(v)}{( {\partial {\cal E}} / {\partial v} )_{M,\,S}}\,; 
\eea
note that $M(v)$ is a slowly evolving total BH mass, 
and ${\cal E}$ and ${\cal F}_{\mathrm {eff}}$ are 
the absorption-corrected binding energy and (effective) fluxes.
The different Taylor approximants integrate this system 
of ordinary differential equations differently. 
In each cases, the time-domain waveforms (in the restricted PN approximation) 
are obtained by inserting the resulting phase 
into, e.g., the leading-order GW polarizations $h_{+,\,\times}$~\eqref{PN-hpx}, 
or the dominant $(\ell,\,m) = (2,\,2)$ mode $h_{22}$~\eqref{h22}.

In the spinning, point-particle case (without BH absorption),
these Taylor families of time-domain waveforms are implemented 
as ``Module LALSimInspiralTaylorXX.c''at \url{https://lscsoft.docs.ligo.org/lalsuite/lalsimulation/group___l_a_l_sim_inspiral_taylor_x_x__c.html}
in LALSimulation;
the equations for the point-particle binaries
are recovered by simply replacing ${\cal E}$ and ${\cal F}_{\mathrm {eff}}$
with a ``standard'' expressions of $E$~\eqref{def-E} 
and $F_{\infty}$~\eqref{def-Finf} as well as 
the constant total mass $M$~\cite{lalsuite,Buonanno:2009zt}.

%%%%%%%%%%%%%%%%%%%%%%%%%%%%%%%%%%%%%%%%%%%%%
\subsubsection{TaylorT1}

The TaylorT1 phase, $\phi^{\mathrm {T1}}(t)$ is obtained 
by solving the system of two ordinary differential equations represented by
Eqs.~\eqref{Kepler} and~\eqref{dvdt} 
with respect to time $t$.
One may use $v^{\mathrm {T1}}(t)$ to calculate the GW amplitude.

%%%%%%%%%%%%%%%%%%%%%%%%%%%%%%%%%%%%%%%%%%%%%%%%%%%%%%%%%%
\subsubsection{TaylorT4}

Based on Ref.~\cite{Buonanno:2002fy}, first Taylor expands the ratio 
in the right-hand side of Eq.~\eqref{dvdt} and then truncate at appropriate
PN order before integrating:
\beq\label{dvdt-T4}
\frac{d v^{\mathrm {T4}}}{d t} 
= 
\frac{d v^{\mathrm {T4}}_{\infty}}{d t} 
+
\frac{d v^{\mathrm {T4}}_{\mathrm H}}{d t}\,,
\eeq
where $d v^{\mathrm {T4}}_{\infty}/{d t}$ describes 
the spinning point-particle contribution 
while $d v^{\mathrm {T4}}_{\rm H}/{d t}$ accounts 
for the BH-absorption correction.
Their formal PN structures are 
\begin{equation}\label{dvdtT4-I0}
\frac{d v^{\mathrm {T4}}_{\infty}}{d t} 
=
\frac{32}{5} \frac{\eta}{M} v^9 
\left\{
{\dot v}^{\mathrm {T4}}_{\mathrm {NS}}
+
{v^3} {\dot v}^{\mathrm {T4}}_{\mathrm {SO}}
+
{v^4} {\dot v}^{\mathrm {T4}}_{\mathrm {SS}}
+
{v^7} {\dot v}^{\mathrm {T4}}_{\mathrm {SSS}}
+ 
O(v^8)
\right\}\,,
\end{equation}
and
\bea\label{dXdt-T4H}
\frac{d v^{\mathrm {T4}}_{\mathrm {H}}}{d t} 
= 
\frac{32}{5} \frac{\eta}{M}  v^{14} 
\left\{
{\dot v}^{\mathrm {T4}}_{{\mathrm {Flux}},5}
+
v^2 
\left(
{\dot v}^{\mathrm {T4}}_{{\mathrm {Flux}},7}
+
\eta\, {\dot v}^{\mathrm {T4}}_{{\mathrm {BH}},7}
\right)
+ 
O(v^3)
\right\}\,.
\eea
One then integrates two ordinary differential equations~\eqref{Kepler} 
and ~\eqref{dvdt-T4} at the same time. The resulting solutions $v^{\mathrm {T4}}(t)$ 
and $\phi^{\mathrm {T4}}(t)$ are TaylorT4 approximants.
We note that among Taylor-based approximants, 
this waveform in the case of non-spinning quasicircular orbits, is the one 
that agrees more with NR waveforms for moderate values 
of the mass ratio $0.5 \lesssim q \leq 1$ ~\cite{Boyle:2007ft}.

%%%%%%%%%%%%%%%%%%%%%%%%%%%%%%%%%%%%%%%%%%%%%%%%%%%%%%%%%%%%%%%%%%
\subsubsection{TaylorT2}

We rewrite Eqs.~\eqref{Kepler} and~\eqref{dvdt} as 
\beq
\frac{d \phi}{d v} 
= 
\frac{v^3}{M(v)} \frac{d t}{d v}\,,
\quad 
\frac{d t}{d v} 
= 
-\frac{( {\partial {\cal E}} / {\partial v} )_{M,\,S}}
{{\cal F}_{\mathrm {eff}}(v)} \,.
\label{T2-t}
\eeq
One can then analytically integrate this system with respect to $v$ 
after re-expanding the right-hand sides of these expressions in PN series. 
The resulting solutions are TaylorT2 phase $\phi^{\mathrm {T2}}(v) $ and time $t^{\mathrm {T2}}(v)$.
We express them as 
\beq
\phi^{\mathrm {T2}}(v) 
= 
\phi^{\mathrm {T2}}_{\mathrm {ref}}
+
\phi^{\mathrm {T2}}_{\infty}(v) 
+
\phi^{\mathrm {T2}}_{\mathrm H}(v)\,,
%\label{phi-T2}
\quad
t^{\mathrm {T2}}(v) 
= 
t^{\mathrm {T2}}_{\mathrm {ref}}
+
t^{\mathrm {T2}}_{\infty}(v) 
+
t^{\mathrm {T2}}_{\mathrm H}(v)\,,
\label{t-T2}
\eeq
where
$\phi^{\mathrm {T2}}_{\mathrm {ref}}$ and $t^{\mathrm {T2}}_{\mathrm {ref}}$
denote the reference phase and time as integration constants.
The spinning point-particle contributions give rise to 
\bea\label{T2-inf}
\phi^{\mathrm {T2}}_{\infty} 
&=& 
-\frac{1}{32 \eta v^5}
\left\{
\phi^{\mathrm {T2}}_{\mathrm {NS}}
+
{v^3} \phi^{\mathrm {T2}}_{\mathrm {SO}}
+
{v^4} \phi^{\mathrm {T2}}_{\mathrm {SS}}
+
{v^7} \phi^{\mathrm {T2}}_{\mathrm {SSS}}
+ 
O(v^8)
\right\}\,, \cr
%%%%%%%%%%%%%%%%%%%%%%%%%%%%%%
t^{\mathrm {T2}}_{\infty}
&=& 
-\frac{5 M}{256 \eta v^8}
\left\{
t^{\mathrm {T2}}_{\mathrm {NS}}
+
{v^3} t^{\mathrm {T2}}_{\mathrm {SO}}
+
{v^4} t^{\mathrm {T2}}_{\mathrm {SS}}
+
{v^7} t^{\mathrm {T2}}_{\mathrm {SSS}}
+ 
O(v^8)
\right\}\,,
\eea
while the BH-absorption corrections are
\bea\label{T2-H}
\phi^{\mathrm {T2}}_{\mathrm {H}} 
&=& 
-\frac{1}{32 \eta }
\left\{
\ln \left( v \right) 
\phi^{\mathrm {T2}}_{\mathrm {Flux},5}
+
{v^2} 
\left( 
\phi^{\mathrm {T2}}_{\mathrm {Flux},7} 
+
\eta \, \phi^{\mathrm {T2}}_{\mathrm {BH},7}
\right)
+
O(v^3)
\right\}\,, \cr
%%%%%%%%%%%%%%%%%%%%%%%%%%%%%%%%%%%%%%%%%%%%%%%%
t^{\mathrm {T2}}_{\mathrm {H}}
&=& 
-\frac{5 M}{256 \eta v^3}
\left\{
t^{\mathrm {T2}}_{\mathrm {Flux},5}
+
{v^2} 
\left( 
t^{\mathrm {T2}}_{\mathrm {Flux},7} 
+
\eta \, t^{\mathrm {T2}}_{\mathrm {BH},7}
\right)
+
O(v^3)
\right\}\,,
\eea
Here, all the terms are expressed in the PN series (analytically). 
TaylorT2 approximants are useful to understand the contributions of each term  
because the total expressions are summed up.

%%%%%%%%%%%%%%%%%%%%%%%%%%%%%%%%%%%%%%%%%%%%%%%%%%%%%
\subsubsection{TaylorT3}

Inverting the PN series $t(v)$ (like TaylorT2 time $t^{\mathrm {T2}}(v)$) 
to obtain $v(t)$ analytically,  
we can re-express the TaylorT2 phase with respect to $t$  
via the relation $\phi(t) \equiv \phi(v(t))$. 
These are the TaylorT3 approximants. 
With the dimensionless time variable 
($t^{\mathrm {T2}}_{\mathrm {ref}}$ is analogue to the coalescence $t_c$
introduced in Eq.~\eqref{T-of-coalescence}), 
\beq
\theta 
\equiv 
\left\{
\frac{\eta}{5M} (t^{\mathrm {T2}}_{\mathrm {ref}} - t)
\right\}^{-1/8}\,, 
\label{def-theta}
\eeq
one obtains
\beq
%\label{phi-T3}
\phi^{\mathrm {T3}}(\theta) 
=
\phi^{\mathrm {T3}}_{\mathrm {ref}}
+
\phi^{\mathrm {T3}}_{\infty}(\theta) 
+
\phi^{\mathrm {T3}}_{\mathrm H}(\theta)\,, 
\quad
%%%%%%%%%%%%%%%%%%%%%%%%%%%%%%
%\label{f-T3}
F^{\mathrm {T3}}(\theta) 
= 
F^{\mathrm {T3}}_{\infty}(\theta) 
+
F^{\mathrm {T3}}_{\mathrm H}(\theta)\,,
\eeq
where $F \equiv (2 d \phi / dt)/(2 \pi) = {v^3}/{(\pi m)}$ is 
the GW frequency of the dominant $(\ell,\, m) = (2,\, 2)$ GW mode, 
and the meaning of the labels ``$\infty$'' and ``H'' are the same as Eq.~\eqref{t-T2}. 
%In Eqs.~\eqref{phi-T3} and~\eqref{f-T3}, 
Their formal PN structures are 
\bea\label{T3-inf}
\phi^{\mathrm {T3}}_{\infty} 
&=& 
-\frac{1}{\eta \theta^5}
\left\{
\phi^{\mathrm {T3}}_{\mathrm {NS}}
+
{\theta^3} \phi^{\mathrm {T3}}_{\mathrm {SO}}
+
{\theta^4} \phi^{\mathrm {T3}}_{\mathrm {SS}}
+
{\theta^7} \phi^{\mathrm {T3}}_{\mathrm {SSS}}
+ 
O(\theta^8)
\right\}\,, \cr
%%%%%%%%%%%%%%%%%%%%%%%%%%%%%%
F^{\mathrm {T3}}_{\infty}
&=& 
\frac{\theta^3}{8 \pi M }
\left\{
F^{\mathrm {T3}}_{\mathrm {NS}}
+
{\theta^3} F^{\mathrm {T3}}_{\mathrm {SO}}
+
{\theta^4} F^{\mathrm {T3}}_{\mathrm {SS}}
+
{\theta^7} F^{\mathrm {T3}}_{\mathrm {SSS}}
+ 
O(\theta^8)
\right\}\,,
\eea
and
\bea\label{T3-H}
\phi^{\mathrm {T3}}_{\mathrm {H}} 
&=& 
-\frac{1}{\eta}
\left\{
\ln \left(\theta \right) 
\phi^{\mathrm {T3}}_{\mathrm {Flux},5}
+
{\theta^2} 
\left(
\phi^{\mathrm {T3}}_{\mathrm {Flux},7}
+
\eta \, \phi^{\mathrm {T3}}_{\mathrm {BH},7}
\right)
+
O(\theta^3)
\right\}\,, \cr
%%%%%%%%%%%%%%%%%%%%%%%%%%%%%%
F^{\mathrm {T3}}_{\mathrm {H}}
&=& 
\frac{\theta^8}{8 \pi M}
\left\{
F^{\mathrm {T3}}_{\mathrm {Flux},5}
+
{\theta^2} 
\left(
F^{\mathrm {T3}}_{\mathrm {Flux},7}
+
\eta \, F^{\mathrm {T3}}_{\mathrm {BH},7}
\right)
+
O(\theta^3)
\right\}\,.
\eea
The TaylorT3 waveform is useful when plotting inspiral GW waveforms in the time domain 
because the GW phase (by using $\phi^{\mathrm {T3}}$) and amplitude (by using $F^{\mathrm {T3}}$)
are written with respect to $t$ directly.

%%%%%%%%%%%%%%%%%%%%%%%%%%%%%%%%%%%%%%%%%%%%%%%%%%%%%%
\subsubsection{TaylorT5}

A variant of the TaylorT2 construction has been adopted to define 
the TaylorT5 approximants in Ref.~\cite{Ajith:2011ec}.
It consists in Taylor expanding in $v$ the right-hand side of the second equation
of Eq.~\eqref{T2-t}, the one for $dt/dv$, 
truncating it to the appropriate order, then taking its inverse, 
and integrating it to obtain $v(t)$.

The phasing is then obtained by substituting $v(t)$ inside the analytical
expression of $\phi(v)$ and direct integration of the Taylor expanded
$d\phi/dv$.

%%%%%%%%%%%%%%%%%%%%%%%%%%%%%%%%%%%%%%%%%%%%%%%%%%%%%%%%%%%
\subsection{Taylor frequency-domain approximants}
%%%%%%%%%%%%%%%%%%%%%%%%%%%%%%%%%%%%%%%%%%%%%%%%%%%%%%%%%%

For an adiabatic BBH inspiral, the frequency-domain waveform 
is conveniently constructed using the stationary phase approximation (SPA), 
described in the subsection ``Stationary phase approximation'' 
(see also, e.g., Ref.~\cite{Mishra:2016whh} and references therein). 
In the case of the dominant $(\ell,\, m) = (2,\, 2)$ GW modes $h_{22}$, 
for instance, we model the Fourier amplitude $A(f)$ and 
the Fourier-domain phase $\Psi(f)$ defined by 
%(recall ${\tilde h}_{2,\,2}(f) = {\tilde h}_{2,\,-2}^{\ast}(-f)$)
\beq\label{h-F}
{\tilde h}_{22}(f)
=
A(f) \, e^{ i ( {\Psi_{\mathrm {SPA}}}(f) - \pi / 4) } \,,
\quad
\Psi_{\mathrm {SPA}}(f) = 2 \pi f t(f) - \Psi(f) \,.
\eeq
The SPA Fourier amplitude with evolving reduced mass $\mu(v)$ is 
\beq
A(f) \simeq 
\left.
\frac{8 \mu(v)} {r} \sqrt{\frac{\pi}{5}} v^2
\left( \frac{3 v^2}{\pi M} \frac{dv}{dt} \right)^{-1/2} \right|_{v = v_f}\,,
\label{def-A}
\eeq
where $v_f \equiv (\pi M f)^{1/3}$ is a dimensionless Fourier parameter 
(normalised by the initial values of the total mass $M = M_I$) 
and $dv/dt$ is given in Eq.~\eqref{dvdt}. 
At the same time, the SPA Fourier-domain phase is obtained 
by solving the set of ordinary differential equations: 
\begin{equation}\label{dpsidt-F2}
\frac{d \Psi_{\mathrm {SPA}}}{df} - 2 \pi t = 0\,, 
\quad
\frac{dt}{df} 
+ 
\frac{\pi M}{3 v^2} 
\frac{( {\partial {\cal E}} / {\partial v} )_{m,\,S}}
{{\cal F}_{\mathrm {eff}}(v)}
= 0\,.
\end{equation}
Again, ${\cal E}$ and ${\cal F}_{\mathrm {eff}}$ are 
the absorption-corrected binding energy and (effective) fluxes: 
recall Eq.~\eqref{balanceH-0}.
The different Taylor approximants integrate Eqs.~\eqref{def-A} and~\eqref{dpsidt-F2} differently. 

In the spinning, nonprecessing point-particle case (without BH absorption)
TaylorF2 waveforms are implemented as ``Module LALSimInspiralTaylorF2.c''at \url{https://lscsoft.docs.ligo.org/lalsuite/lalsimulation/group___l_a_l_sim_inspiral_taylor_x_x__c.html}
in LALSimulation; 
the equations for the point-particle binaries are recovered 
by replacing ${\cal E}$ and ${\cal F}_{\mathrm {eff}}$ in Eq.~\eqref{dpsidt-F2} 
with $E$~\eqref{def-E}, $F_{\infty}$~\eqref{def-Finf} 
and the constant total mass $M$~\cite{lalsuite,Buonanno:2009zt}.

%%%%%%%%%%%%%%%%%%%%%%%%%%%%%%%%%%%%%%%%%%%%%%%%%%%%%%%%%%%%%%%%%%%%
\subsubsection{TaylorF1}

The direct numerical integration of Eqs.~\eqref{def-A} and~\eqref{dpsidt-F2} 
gives the TaylorF1 amplitude $A^{\mathrm {F1}}(f)$ and 
the TaylorF1 phase $\Psi_{\mathrm {SPA}}^{\mathrm {F1}}(f)$, respectively. 
One may use the TaylorT4 $dv^{\rm T4}/dt$~\eqref{dvdt-T4} 
as input PN expression for $dv/dt$ when solving Eq.~\eqref{def-A}.

%%%%%%%%%%%%%%%%%%%%%%%%%%%%%%%%%%%%%%%%%%%%%%%%%%%%%%%%%%%%%%%%%%%%
\subsubsection{TaylorF2}

Drawing an analogy with TaylorT2, 
we re-expand the right-hand sides of Eqs.~\eqref{def-A} and~\eqref{dpsidt-F2} 
in the PN series and analytically integrate them with respect to $f$. 
The TaylorF2 phase is then given by 
\begin{equation}\label{phi-F2}
\Psi^{\mathrm {F2}}_{\mathrm {SPA}}(f)
= 
2 \pi f t_c - \Psi_c 
+
\Psi^{\mathrm {F2}}_{\infty}(f)
+
\Psi^{\mathrm {F2}}_{\mathrm {H}}(f)\,,
\end{equation}
where $t_c$ and $\Phi_c$ are constants that we can choose arbitrarily.
The spinning point-particle contribution is
\beq\label{F2-inf}
\Psi^{\mathrm {F2}}_{\infty} (f)
= 
\frac{3}{128 \eta v^5}
\left\{
\Psi^{\mathrm {F2}}_{\mathrm {NS}}
+ 
{v_f^3} \Psi^{\mathrm {F2}}_{\mathrm {SO}} 
+
{v_f^4} \Psi^{\mathrm {F2}}_{\mathrm {SS}} 
+
{v_f^7} \Psi^{\mathrm {F2}}_{\mathrm {SSS}} 
+ 
O(v_f^8)
\right\}\,,
\eeq
%(see, e.g, Ref.~\cite{Boyle:2007ft} for explicit expressions)
while the BH-absorption contributions are  
%of the horizon fluxes 
%and the secular change in BH masses and spins are 
\bea\label{F2-H}
&& \Psi^{\mathrm {F2}}_{\mathrm {H}}(f) 
=  \cr
&& \quad 
\frac{3}{128 \eta}
\left[
\left\{
1 + 3 \ln \left( \frac{v_f}{v_{\mathrm {reg}}} \right) 
\right\}
\Psi^{\mathrm {F2}}_{\mathrm {Flux},5}
+
{v_f^2} \left(
\Psi^{\mathrm {F2}}_{\mathrm {Flux},7}
+
\eta \, \Psi^{\mathrm {F2}}_{\mathrm {BH},7}
\right)
+
O(v_f^3)
\right]\,,
\eea
where the constant $v_{\rm {reg}}$ can be chosen arbitrary.

The TaylorF2 amplitude $A^{\rm F2}(f)$ is given by inserting 
TaylorT4 $dv^{\rm T4}/dt$~\eqref{dvdt-T4} into Eq.~\eqref{def-A} 
and re-expanding it in the PN series and truncate at $3.5$PN order. 
The most commonly used TaylorT2 amplitude is, however, in the restricted PN approximation given by 
\beq
A^{\rm F2} (f) \simeq 
\frac{{\cal M}^{5/6}}{r}\,\sqrt{\frac{2}{3 \pi^{1/3}}}\,f^{-7/6} 
\left\{1 + O(v^2) \right\}\,.
\label{A-F2}
\eeq

It is worth noting that the spin effects at the leading $1.5$PN order 
(i.e., the leading spin-orbit term) in the SPA amplitude and phase 
are encoded in a single spin parameter,
the effective aligned spin~\cite{Damour:2001tu}, 
$\chi_{\rm eff} \equiv (m_1 \chi_1 + m_2 \chi_2) /M$ 
or the reduced spin parameter~\cite{Ajith:2011ec}
\beq
\chi_{\rm PN} \equiv 
\chi_{\rm eff} - \frac{38\, \eta}{113} (\chi_1 + \chi_2)\,,
\label{def-chiPN}
\eeq
used for waveform calibration against NR waveform  in inspiral-merger-ringdown waveforms
which we will introduce in the next section.

%%We note that when we plot the frequency domain waveform
%%with detector's sensitivity curves in $1/\sqrt{\rm Hz}$, 
%%which are the square root of the power spectral density
%%of the detector noise $S(f)$,
%%we may use~\cite{Moore:2014lga}
%%\beq
%%2\, f^{1/2} |\tilde{h}(f)| \,,
%%\eeq
%%We should note that the plot with the above quantity
%%is different from
%%that with the characteristic strain
%%at \url{http://gwplotter.com/}
%%~\cite{GWplotter}.

%%%%%%%%%%%%%%%%%%%%%%%%%%%%%%%%%%%%%%%%%%%%%%%%%%%%%%%%%%%%%%%%%%%%%%%%%
\subsection{Beyond spinning, nonprecesssing Binary Black Hole cases }
%%%%%%%%%%%%%%%%%%%%%%%%%%%%%%%%%%%%%%%%%%%%%%%%%%%%%%%%%%%%%%%%%%%%%%%%%%

Until this subsection, we have confined our attention to the simplest adiabatic inspiral 
of a spinning, nonprecessing BBH without orbital eccentricity. 
%(implicitly) assuming that the propagation of GW signals across cosmological distances 
%are not affected by the expansion of the Universe. 
The leap from this narrow case to a generic case of binary configurations
-- in particular, one of the component compact objects is a NS rather 
than a BH -- comes with a number of consequences.  

Below we briefly review some of the key elements when constructing PN templates 
for generic inspirals; 
the inclusion of merger and ringdown phase with completely generic waveform
models will be summarized in the next section.

%%%%%%%%%%%%%%%%%%%%%%%%%%%%%%%%%%%%%%%%%%%%%%%%%
{\it Eccentricity}.---
%%%%%%%%%%%%%%%%%%%%%%%%%%%%%%%%%%%%%%%%%%%%%%%%%
%
The radiative loss of the orbital energy and angular momentum to GWs 
circularizes the orbits of inspirals.
The averaged rates of change of orbital eccentricity $e (\leq 1)$ 
due to the radiative losses in the adiabatic approximation are estimated as 
(see, e.g., Chapter~12 of Poisson and
Will~\cite{2014grav.book.....P}, and Ref.~\cite{Peters:1964zz})
\beq
\frac{de}{dt}
=
-\frac{304}{15}\, \eta\,\frac{e}{a} \left( \frac{M}{a} \right)^3\,
(1 - e^2)^{-5/2}\, \left( 1 + \frac{121}{304} e^2 \right)\,,
\label{dote}
\eeq
where $a$ is the semi-major axis of the ellipse (not the Kerr spin parameter here)
and we assume Newtonian elliptic orbits for simplicity. 
That is, the eccentricity always decreases when the orbit shrinks as approaching the merger phase. 
In fact, assume a small eccentricity limit $e \ll 1$, Eq.~\eqref{dote}, can be integrated 
(with the help of $da/dt$ etc.) to give $e \simeq e_{\rm ini} ({a}/{a_{\rm ini}})^{19/12}$, 
where $a_{\rm ini}$ and $e_{\rm ini}$ are initial values. 
We clearly see that the orbit is circularised quite fast 
toward the late inspiral phase. 

Nevertheless, there is emerging need for GW models for the quasieccentric inspirals 
because it will be an important source for the next-generation GW detectors 
both on the ground (e.g., KAGRA+~\cite{Michimura:2019cvl}, Voyager (\url{https://dcc.ligo.org/LIGO-G1602258/public}), 
Einstein Telescope~\cite{Hild:2008ng}, Cosmic Explorer~\cite{Reitze:2019iox}, etc) 
and in the space (e.g., LISA~\cite{Audley:2017drz}, (B-)DECIGO~\cite{Kawamura:2020pcg}, 
TianQin~\cite{Mei:2020lrl}, etc), 
which have wider sensitivity band for the early inspiral phase; 
see also Chapter~3 by Gair et al. and Chapter~7 by Lueck et al. in this book. 
There is an ongoing program of development on the waveforms 
for the quasieccentric binaries. 
%As for the eccentricity effects, for example, 
A theoretical GW waveform for non-spinning eccentric binaries
has been developed 
in Refs.~\cite{Huerta:2014eca,Moore:2016qxz,Tanay:2016zog,Tiwari:2019jtz,Tiwari:2020hsu} 
with the quasi-Keplerian formalism
(see also Chapter~10 of Ref.~\cite{Blanchet:2013haa} and references therein).
This waveform is valid for the small initial orbital eccentricity,  
%$e_{\rm ini} \lesssim 0.1$, 
and its frequency-domain model is implemented in LALSimulation as TaylorF2Ecc at \url{https://lscsoft.docs.ligo.org/lalsuite/lalsimulation/group___l_a_l_sim_inspiral_taylor_f2_ecc__c.html} 
(see also EccentricTD/FD modules). 
%~\cite{lal_i_t_f2_ecc}.
However, a GW waveform that performs well even at large eccentricities
requires further development. Recent efforts can be found 
in, e.g., Refs.~\cite{Moore:2018kvz,Moore:2019xkm} and Chapter~33 
by Loutrel in this book.

%Therefore, in a search for eccentric BBH mergers
%shown by Ref.~\cite{Salemi:2019owp},
%the coherent WaveBurst (cWB) algorithm has been used. 
%This method does not rely on theoretical GW waveforms
%for binaries.

%%%%%%%%%%%%%%%%%%%%%%%%%%%%%%%%%%%%%%%%%%%%%%%%%
{\it Spin precession}.---
%%%%%%%%%%%%%%%%%%%%%%%%%%%%%%%%%%%%%%%%%%%%%%%%%
In the nonprecessing system, the spins of the compact objects are (anti-)parallel 
to the orbital-angular momentum; the orbital place of binary is fixed. 
If the spins are not aligned with the orbital angular momentum 
the spin-orbit and spin-spin effects lead to precession of the spins and orbital plane, while the total angular momentum is constant, modulo
angular momentum loss via radiation.
The precession timescale for the spinning binaries is 
\beq
t_{\rm P} 
=
\frac{|{\bf S}_a|}{|d{\bf S}_a/dt|}
\sim M v^{-5}\,,
\label{t_P}
\eeq
which is shorter than the radiation reaction timescale 
$t_{\rm RR}/M \propto v^{-8}$ given in Eq.~\eqref{t_RR}, hence neglecting
losses the orbital and spin angular momentum precess 
around the total angular momentum. 
Schematically, the (orbital-averaged) spin precession equations 
that govern the conservative evolution of spin vectors are 
(see, e.g., Chapter~9 of Poisson and Will~\cite{2014grav.book.....P} 
and Ref.~\cite{Buonanno:2005xu} for the explicit expressions) 
\bea
\frac{d {\bf S}_a}{dt} 
= \left(\bm{\Omega}_a^{\rm SO} + \bm{\Omega}_a^{\rm SS} \right) \times {\bf S}_a \,, 
\label{eq:PNEOM_spin}
\eea
where $\bm{\Omega}_a^{\rm {SO,\,SS}}$ are spin-orbit and spin-spin pieces 
of the precessional angular velocity. 
The frequency $\bm{\Omega}_a^{\rm {SO}}$ describes the spin-orbit (geodetic) precession 
of the spin vectors while $\bm{\Omega}_a^{\rm {SS}}$ is responsible 
for the spin-spin (frame-dragging) precession of those spin vectors.

The precession determined by Eq.~\eqref{eq:PNEOM_spin} then modulates the waveform, 
adding rich periodic structure, as the angle between the normal to the orbit
and the line of sight also precesses. 
As a result, the GW mode $(\ell,\,m) = (2,\,2)$ is no longer guaranteed to be the ``dominant'' one; 
all the $\ell = 2$ modes become relevant. 
These complications make the modeling of precessing waveform challenging, 
but they gain us access to more binary parameters
that can be hard to measure in nonprecessing binaries, due to their degeneracies with other parameters that can be disentangled by observing the source's plane
from different angles during the very same inspiral.

For more recent effort to develop the precessing binary waveforms, 
see, e.g., a concise review by Hannam~\cite{Hannam:2013pra} (and references therein, also see \cite{Chatziioannou:2017tdw}). 
Various spin precessing Taylor waveforms are implemented in LALsimulation and are found 
at \url{https://lscsoft.docs.ligo.org/lalsuite/lalsimulation/group___l_a_l_sim_inspiral_spin_taylor__c.html}.

%%%%%%%%%%%%%%%%%%%%%%%%%%%%%%%%%%%%%%%%%%%%%%%%%
{\it Matter effects}.---
%%%%%%%%%%%%%%%%%%%%%%%%%%%%%%%%%%%%%%%%%%%%%%%%%
When one of binary components is not a BH, 
but a material body as NS, 
it can acquire a quadrupole moment induced by, e.g., the tidal field
generated by its companion.  
It is particularly interesting since
the deformation giving rise to the tidally induced quadrupole 
bears the imprint of the equation of state of the NS -- i.e., 
the microscopic property of strongly interacted nuclear matter  
under extreme conditions of pressure and density.

The induced quadrupole moment of NSs is mainly characterised by two physical effects. 
The first one is due to (static) tides,
as non-spinning NS of mass $m_a$ and radius $R_a^{\rm NS} \ll R$ 
in a binary is subjected to a tidal field by the companion. 
In a Newtonian gravity (for simplicity), 
a tidally induced quadrupole moment is given by 
\beq
Q_a^{\rm tidal} \sim \lambda_a \frac{m_b}{R^3}\,,
\eeq
where $\lambda_a =k_a (R_a^{\rm NS})^5$ is the (static) tidal deformability 
defined in terms of the dimensionless (gravitational) Love number $k$ 
that depends on the NS's equation of state.
A measurement of $k_a$ or $\lambda_a$ through the use of GW therefore provides 
unique insight into NS matter; 
we note that $k_a = 0$ for a non-spinning, Schwarzschild BH 
(the tidally induced quadrupole moments of Schwarzschild BH 
vanish in the static case~\cite{Binnington:2009bb,Damour:2009va,Kol:2011vg} 
and is proportional to the inducing field time
derivatives, being a dissipative effect, giving rise to the horizon fluxes 
discussed in the previous section~\cite{Poisson:2004cw}).
The GW astronomy on the NS tidal deformability is 
described in detail, e.g., 
by Chatziioannou~\cite{Chatziioannou:2020pqz}; 
see also Chapter~11 by Baiotti and Chapter~14 by Foucart in this book.

For non-spinning BNSs, the family of PN templates with tidal interactions is nicely summarised  
in, e.g., Ref.~\cite{Narikawa:2019xng} and Chapter~4 by Dietrich et al.~\cite{Dietrich:2020eud}.
%(see also, e.g., Ref.~\cite{} for the tidal contributions in spinning BNSs)
As an example, we briefly summarise the \emph{Kyoto}'s phenomenological model 
by Kawaguchi et al.~\cite{Kawaguchi:2018gvj}, 
which combines the PN tidal corrections with phenomenological terms obtained
by fits to Kyoto's high-precision NR waveforms.
In this model, TaylorF2 (strain) $\tilde{h}(f) = A(f) \,e^{i\Psi(f)}$ 
with the tidal contributions may take the split form of 
\beq
A(f) = A_{\rm PP}(f) + A_{\rm Tidal}(f) \,,
\quad
\Psi(f) = \Psi_{\rm PP}(f) + \Psi_{\rm Tidal}(f) \,,
\eeq
where  ``PP'' and ``tidal''are individual contributions 
from the non-spinning point particle and tidal interactions, respectively.
%for the GW amplitude, and $\Psi_{\rm Tidal}$we introduce a tidal term $A_{\rm Tidal}$
%for the GW phase. Here, %$5$PN effects: 
The tidal terms $A_{\rm Tidal}(f)$ and $\Psi_{\rm Tidal}(f)$ 
are then given by (note that $x\equiv v^2$)
\bea
%%%%%%%%%%%%%%%%%%%%%%%%%%%%%%%%%%%%%%%%%%%%%%%%%%%%%%%%%%%%%%%%%%%%%%%%%%%%%%%%%%%%%%%%%%%
A_{\rm Tidal}^{\rm KyotoTidal}
&=& \sqrt{\frac{5 \pi \eta}{24}} \frac{M^2}{r} \tilde{\Lambda} x^{-7/4}
\times \left( - \frac{27}{16} x^5 - \frac{449}{64} x^6 - b x^{r} \right) \,, \cr 
%%%%%%%%%%%%%%%%%%%%%%%%%%%%%%%%%%%%%%%%%%%%%%%%%%%%%%%%%%%%%%%%%%%%%%%%%%%%%%%%%%%%%
\Psi_{\rm Tidal}^{\rm KyotoTidal}
&=& \frac{3}{128\eta} \left\{ -\frac{39}{2} \tilde{\Lambda} 
\left( 1+a \tilde{\Lambda}^{2/3} x^p \right) \right\} x^{5/2}
\cr
&& \times \left( 
1 + \frac{3115}{1248} x - \pi x^{3/2} + \frac{28024205}{3302208} x^2 
- \frac{4283}{1092} \pi x^{5/2} \right) \,,
\label{KyotoTidal}
\eea
with $a=12.55$, $p=4.240$, $b=4251$ and $r=7.890$. 
Here, $\tilde{\Lambda}$ is the combination of the individual 
(dimensionless) tidal deformability $\Lambda_a \equiv \lambda_a / m_a^5$
defined by~\cite{Flanagan:2007ix}
\bea
\tilde{\Lambda} = \frac{16}{13} \frac{(m_1+12m_2)m_1^4\Lambda_1+(m_2+12m_1)m_2^4\Lambda_2}
{M^5} \,. 
\label{def-tildeLambda}
\eea
This parameter characterises the leading-order (relative $5$PN) tidal effects in the waveforms.
%and used by measuring LIGO-Virgo BNS events, GW170817 and GW 190425. 

TaylorF2 with tidal effects was used to analyze the first BNS event, GW170817~\cite{TheLIGOScientific:2017qsa}, 
and it is implemented in LALSimulation as a part of 
``Module LALSimInspiralTaylorXX.c''at \url{https://lscsoft.docs.ligo.org/lalsuite/lalsimulation/group___l_a_l_sim_inspiral_taylor_x_x__c.html}. 
For more improved analysis of GW170817~\cite{Abbott:2018wiz},
a different model was used as a reference.
In GWTC-1~\cite{LIGOScientific:2018mvr}, 
this is also used as a frequency-domain model for GW170817.
For example, we will see the {\rm SEOBNRv4\_ROM\_NRTidal} and
{\rm IMRPhenomPv2\_NRTidal} models in the next section.

%%%%%%%%%%%%%%%%%%%%%%%%%%%%%%%%%%%%%%%%%%%%%%%%%%%%%%%%%%%%%%%%%%%%%

The second effect that gives rise to the induced quadrupole moment is the rotation. 
The spinning motion deforms the NS by creating a distortion in its mass distribution, 
which also depends on the NS's equation of state~\cite{Laarakkers:1997hb}. 
The resultant (relativistic) spin-induced quadrupole moment 
is given by~\cite{Poisson:1997ha}
\beq
Q_a^{\rm spin} \simeq  \alpha \,{\chi}_a^2 \,m_a^3 \,,
\eeq
where $\alpha$ is the (dimensionless) spin deformability. 
Although the known NS has low spin in general 
(${\chi}_a \sim 0.2$ or less~\cite{TheLIGOScientific:2017qsa,Abbott:2020uma}), 
in principle, the measurement of $\alpha$ for highly spinning BNSs would 
provide another GW probe to NS matter~\cite{Harry:2018hke}. 
We note that $\alpha=1$ for Kerr BHs, followed by its well-known ``no-hair'' property.

By likewise writing $\Psi_{\rm Tidal}$, the correction to the TaylorF2 phase 
due to the NS's quadrupole spin-deformation is given by~\cite{Poisson:1997ha,Krishnendu:2017shb}. 
\beq
\Psi_{QM}
=
\frac{3}{128\eta} \left(
-25 {\tilde {Q}}
\right) v^{-1} + O(v) \,,
\eeq
where ${\tilde Q}$ is a certain combination of the individual spin-induced quadrupole 
deformation $Q_a^{\rm spin}$ and spins ${\chi}_a$, 
which characterise the leading-order (relative $2$PN) effects in the waveform.
It is implemented in LALSimulation within ``Module LALSimInspiralSpinTaylor.c''
(\url{https://lscsoft.docs.ligo.org/lalsuite/lalsimulation/group___l_a_l_sim_inspiral_spin_taylor__c.html}), 
and the quadrupole spin-deformation contribution to energy, flux, and phasing
terms are coded in \url{https://lscsoft.docs.ligo.org/lalsuite/lalsimulation/_l_a_l_sim_inspiral_p_n_coefficients_8c.html}.

%%%%%%%%%%%%%%%%%%%%%%%%%%%%%%%%%%%%%%%%%%%%%%%%%%%%%%%%%%%%%%%%%%%%%%%%%%%%%%%%%
\section{Full Inspiral-Merger-Ringdown waveform models}
%%%%%%%%%%%%%%%%%%%%%%%%%%%%%%%%%%%%%%%%%%%%%%%%%%%%%%%%%%%%%%%%%%%%%%%%%%%%%%%%%%

In our discussion so far, we have focused exclusively on 
the GW waveforms from adiabatic inspirals within the PN approximation. 
However, this is just a part of GW signals from the coalescence of compact-object binaries 
that can be observed by LIGO, Virgo, and KAGRA. 

As the separation of binary shrinks (by ``chirping'' the frequency $v \to 1$), 
the binary dynamics moves on to the merger phase, 
and the PN calculations become more and more inaccurate; 
recall our discussion on the accuracy of the PN approximants. 
The modelling in the late inspiral phase should be modified to enhance 
the accuracy of PN approximation. 
The transition frequency from inspiral to the merger phase is roughly estimated 
by the GW frequency at ISCO of Schwarzschild BH with the mass $M_{\rm Sch}$: 
\beq
f_{\rm ISCO} \approx 73.28 \left( \frac{M_{\rm Sch}}{60 M_{\odot}} \right)^{-1}
\,{\rm Hz}
\approx 1570 \left( \frac{M_{\rm Sch}}{2.8 M_{\odot}} \right)^{-1}
\,{\rm Hz} \,.
\eeq
This is well in the sensitive frequency band of LIGO, Virgo and KAGRA.

Furthermore, the ringdown phase followed by the merger phase has 
the typical GW frequency at~\cite{Berti:2009kk}
\beq
f_{\rm Ring} 
= 538.4 \left( \frac{M_{\rm Rem}}{60 M_{\odot}} \right)^{-1}
\left\{ 1.5251 - 1.1568 (1-\alpha_{\rm Rem})^{0.1292} \right\}
\,{\rm Hz} \,,
\eeq
where $M_{\rm Rem}$ and $\alpha_{\rm Rem}$ are
the mass and non-dimensional spin of the remnant Kerr BH after merger.
The above expression gives 198.3 Hz 
for $M_{\rm Rem}=60M_{\odot}$ and $\alpha_{\rm Rem}=0$.
This ringdown frequency, especially for BBHs, is again well 
in the sensitivity band of the Advanced LIGO, Virgo, and KAGRA. 
Therefore, it is indispensable to have a ``full'' waveform, 
including the merger and ringdown phases in addition to the PN model 
for the inspiral phase, in order to maximise our ability to GW data analysis.

The goal of this section is to briefly survey such a full, inspiral-merger-ringdown (IMR) waveform 
actually implemented and used for the GW data analysis of the first, second 
and third observing runs (O1, O2, and O3) of Advanced LIGO and Virgo; 
more details are covered in Chapter~35 by McWilliams in this book, 
and the IMR waveform models used in the up-to-date GWTC-2~\cite{Abbott:2020niy} 
are summarised in Table III of that paper. 
There are two main families of the IMR waveform; 
the effective-one-body (EOB) approach (in the time-domain), 
and the ``phenomenological'' (IMRPhenom) models (in the frequency-domain). 
%While this single section is too short to cover 
All the details of these two methods in LALsimulation can be found 
at \url{https://lscsoft.docs.ligo.org/lalsuite/lalsimulation/group___l_a_l_sim_i_m_r__h.html}.
%in Ref.~\cite{lal_imr}.
We should note, however, that these methods in LALsimulation ignore any of contributions 
of BH absorption discussed in the previous section.

%%%%%%%%%%%%%%%%%%%%%%%%%%%%%%%%%%%%%%%%%%%%%%%%
\subsection{Effective-one-body (EOB) approach}
%%%%%%%%%%%%%%%%%%%%%%%%%%%%%%%%%%%%%%%%%%%%%%%%
An effective-one-body (EOB) approach~\cite{Buonanno:1998gg,Buonanno:2000ef} 
is an analytical framework to cover the full range of the inspiral, merger, and ringdown phases, 
making use of a variety of analytical approximation methods, such as the PN
theory and the Black Hole Perturbation (BHP) theory, and NR data as calibrations. 
A brief review of the EOB approach is given in Refs.~\cite{Damour:2008yg,Damour:2012mv}; 
see also, e.g., Refs.~\cite{Damour:2016gwp,Antonelli:2019fmq} 
(and reference therein) for latest developments.

The starting point of the EOB approach is to precisely describe 
the orbital dynamics of binaries (as a source of GW waveform).
First, one conveniently maps the real two-body PN Hamiltonian (for their relative motion) 
to an ``effective'' test-particle Hamiltonian $H^{\rm eff}$  
of non-geodesic motion in a fictitious effective spacetime, 
so as to construct the so-named EOB (or ``improved real'') Hamiltonian: 
\bea
H^{\rm EOB} \equiv M \sqrt{1 + 2 \eta \left(\frac{H^{\rm eff}}{\mu} - 1\right)} \,.
\eea
In general, $H^{\rm EOB}$ improves the convergence of PN series 
of the original PN Hamiltonian.
Second, the radiation reaction to the system, 
which is another piece to describe the radiative dynamics of binaries, 
is prepared from, e.g., the PN and BHP results of the GW fluxes 
with a resummation such as a Pad\'{e} approximation 
or the factorised resummation~\cite{Damour:2007xr,Damour:2007yf}.
Third, one introduces some adjustable free parameters to the Hamiltonian and fluxes  
and calibrates them against the results of NR simulations 
(and that of the BHP and self-force theory~\cite{Barack:2018yvs} in the small-mass-ratio limit, $q \to 0$).
In particular, the EOB models calibrated to NR simulations are dubbed as ``EOBNR.''

The next step is to construct GW waveforms from the obtained EOB orbital dynamics.
The inspiral-plus-plunge GW waveform $h_{\rm insplunge}$
is derived from the orbital motion, 
based on the improved resummation of PN (multipolar) 
inspiral waveforms~\cite{Damour:2008gu,Pan:2010hz},  
including non-quasicircular effects~\cite{Damour:2007xr,Damour:2007yf}. 
This waveform is connected smoothly to a ringdown GW waveform $h_{\rm ringdown}$
which consists of several quasinormal modes of the remnant BH after merger,
around a matching time $t_{\rm match}$
(see Ref.~\cite{Berti:2018vdi} and references therein
for the ringdown phase).
The full GW waveform is then schematically written as~\cite{Damour:2007yf}
\beq
h_{\rm EOB}(t) = \theta(t_{\rm match}-t)
\,h_{\rm insplunge}(t)
+ \theta(t-t_{\rm match})
\,h_{\rm ringdown}(t) \,,
\eeq
where $\theta(t)$ is the Heaviside-step function.

In Ref.~\cite{Buonanno:2007pf}, the EOBNRv1 model was proposed for non-spinning BBHs.
This was calibrated to NR simulations
with mass ratios, $m_1/m_2 = 1,\, 3/2,\, 2$, and $4$.
In Ref.~\cite{Pan:2011gk},
the EOBNRv2 model was proposed for the same non-spinning case.
This was also calibrated to NR simulations
with mass ratios, $m_1/m_2 = 1,\, 2,\, 3,\, 4$ and $6$.
The above two GW waveform approximants include 
not only the dominant $(\ell ,\,|m| ) = (2,\,2)$ modes, 
but also some subdominant harmonic modes; see also Section~II-a in Ref.~\cite{Abadie:2011kd}.
In the classical GW data analysis of the LIGO fifth science run (S5)~\cite{Abadie:2011kd}, these two EOBNR models were used 
as an IMR theoretical template, 
%(note that Section~II-a in Ref.~\cite{Abadie:2011kd}
%is also useful to briefly see the EOBNRv1/v2 models).
and the EOBNRv2 model is available at \url{https://lscsoft.docs.ligo.org/lalsuite/lalsimulation/group___l_a_l_sim_i_m_r_e_o_b_n_rv2__c.html};
%Ref.~\cite{lal_imr_eobnrv2}.
note, however, that EOBNRv1/v2 models have been superseded by more recent developments 
(such as SEOBNR family below) and they are no longer used in the modern LAL simulation.

\subsubsection{SEOBNR family}
%%%%%%%%%%%%%%%%%%%%%%%%%%%%%%%%%%%%%%%%%%%%%%%%%
{\it SEOBNRv1/v2}.---
%%%%%%%%%%%%%%%%%%%%%%%%%%%%%%%%%%%%%%%%%%%%%%%%%
In Ref.~\cite{Taracchini:2012ig,Taracchini:2013rva} (and references therein),
%Refs.~\cite{Taracchini:2013rva},
EOB models have been presented
for spinning, nonprecessing BBHs 
(the first character, ``S'' in SEOBNR denotes spin).
The SEOBNRv1 and SEOBNRv2 models are available
at \url{https://lscsoft.docs.ligo.org/lalsuite/lalsimulation/group___l_a_l_sim_i_m_r_spin_aligned_e_o_b__c.html}.
%Ref.~\cite{lal_imr_seobnr_aligned}.
%
The parameters of the first BBH event, GW150914~\cite{Abbott:2016blz} 
was evaluated by the SEOBNRv2 model. This was also used
in the detailed study on the properties of
GW150914~\cite{TheLIGOScientific:2016wfe}.
SEOBNRv2\_ROM\_DoubleSpin~\cite{Field:2013cfa}
which speeds up the waveform generation
with reduced-order modelling (ROM) \cite{Purrer:2014fza} was also used.

%\subsubsection{SEOBNRv3}
%%%%%%%%%%%%%%%%%%%%%%%%%%%%%%%%%%%%%%%%%%%%%%%%%
{\it SEOBNRv3}.---
%%%%%%%%%%%%%%%%%%%%%%%%%%%%%%%%%%%%%%%%%%%%%%%%%
In Ref.~\cite{Pan:2013rra}, the SEOBNRv3 model has been presented
as a fully precessing waveform model for BBH coalescence.
The SEOBNRv3 model is available
at \url{https://lscsoft.docs.ligo.org/lalsuite/lalsimulation/_l_a_l_sim_i_m_r_spin_prec_e_o_b_8c.html}.
%Ref.~\cite{lal_imr_seobnrv3}.
%
This SEOBNRv3 model was used in the detailed study on the properties of
GW150914~\cite{TheLIGOScientific:2016wfe}.
Also, in GWTC-1~\cite{LIGOScientific:2018mvr}, 
the SEOBNRv3 model was used to analyse generic two-spin precession dynamics.

%\subsubsection{SEOBNRv4}
%%%%%%%%%%%%%%%%%%%%%%%%%%%%%%%%%%%%%%%%%%%%%%%%%
{\it SEOBNRv4/v4HM}.---
%%%%%%%%%%%%%%%%%%%%%%%%%%%%%%%%%%%%%%%%%%%%%%%%%
For spinning, nonprecessing BBHs, 
the SEOBNRv4 model~\cite{Bohe:2016gbl} is an improvement
of the SEOBNRv2 model with calibration to 141 NR waveforms including the spin effects.
As a further improved version of the SEOBNRv4 model with higher harmonics,
the SEOBNRv4HM model is presented
(HM standing for ``Higher Modes'')~\cite{Cotesta:2018fcv}.
The SEOBNRv4 and SEOBNRv4HM models are available
at \url{https://lscsoft.docs.ligo.org/lalsuite/lalsimulation/group___l_a_l_sim_i_m_r_spin_aligned_e_o_b__c.html}
%Ref.~\cite{lal_imr_seobnr_aligned}
where all aligned spin, i.e., nonprecessing models, are summarized.

For the analysis of GW170817 in Ref.~\cite{Abbott:2018wiz} 
with the help of the NRTidal model for the tidal effects 
(a hybrid model mainly based on the PN tidal phase corrections up to the $7.5$PN order  
and the calibration against the high-precision NR data) 
~\cite{Dietrich:2018uni,Dietrich:2017aum,Dietrich:2019kaq}
$$
{\rm SEOBNRv4\_ROM\_NRTidal} = {\rm SEOBNRv4\_ROM}
+ {\rm NRTidal}
$$
was used as a signal model of BNS mergers.

In GWTC-1~\cite{LIGOScientific:2018mvr}, 
this was also used as a frequency-domain model for GW170817.
In the analysis of GW190814~\cite{Abbott:2020khf},
there was no measurable tidal signature
although the SEOBNRv4\_ROM\_NRTidalv2\_NSBH model
with phenomenological tidal effects and the NS's tidal disruption was applied.
The SEOBNRv4T model~\cite{Hinderer:2016eia}
which gives a time-domain waveform with analytical dynamic tide effects
has been also used in GWTC-1~\cite{LIGOScientific:2018mvr} for GW170817.

%\subsubsection{SEOBNRv4HM/SEOBNRv4PHM}
%%%%%%%%%%%%%%%%%%%%%%%%%%%%%%%%%%%%%%%%%%%%%%%%%
{\it SEOBNRv4P/v4PHM}.---
%%%%%%%%%%%%%%%%%%%%%%%%%%%%%%%%%%%%%%%%%%%%%%%%%
The EOBNR model with higher multipoles for precessing binaries
(see Refs.~\cite{Cotesta:2018fcv,Ossokine:2020kjp} and references therein)
%~\cite{Pan:2013rra, Babak:2016tgq, Ossokine:2020kjp}
was used for GW190412~\cite{LIGOScientific:2020stg}.
Also, this model has been used for GW190814~\cite{Abbott:2020khf}, 
GW190521~\cite{Abbott:2020tfl}, and GWTC-2~\cite{Abbott:2020niy} 
as a BBH waveform model.
%The nonprecessing model is called SEOBNRv4HM, 
The precessing model SEOBNRv4PHM also includes  
its restriction to the dominant GW modes, SEOBNRv4P.

%%%%%%%%%%%%%%%%%%%%%%%%%%%%%%%%%%%%%%%%%%
\subsubsection{TEOBResumS}
%%%%%%%%%%%%%%%%%%%%%%%%%%%%%%%%%%%%%%%%%

The TEOBResumS has been the first to implement a tidal description up to merger
verified with NR simulations \cite{Bernuzzi:2014owa,Akcay:2018yyh,Bernuzzi:2012ci,Nagar:2018zoe}, and it is the only model that implements a binary NS post-merger completion \cite{Breschi:2019srl} so to give a complete description of the signal emitted by  binary NSs.
Note that especially the method of calibration against NR simulations
is different from the above EOBNR family; see Section VI of Ref.~\cite{Nagar:2018zoe}.
Spin interactions in the BNS waveforms are included at next-next-leading order
\cite{Nagar:2018plt} and include precession effects \cite{Akcay:2020qrj}.
The BBH sector implements higher modes \cite{Nagar:2020pcj} and eccentricity \cite{Chiaramello:2020ehz}
and can model hyperbolic mergers \cite{Nagar:2020xsk}.

It is a time-domain waveform that has been used in GWTC-1~\cite{LIGOScientific:2018mvr}
for GW170817 and GWTC-2~\cite{Abbott:2020niy}.  
TEOBResumS is available at \url{https://bitbucket.org/eob_ihes/teobresums/wiki/Home}
%Ref.~\cite{TEOBResumS} 
and included in its non-spinning, reduced-order modeling (ROM) version in the LAL's module
\url{https://lscsoft.docs.ligo.org/lalsuite/lalsimulation/_l_a_l_sim_inspiral_t_e_o_b_resum_r_o_m_8c.html}.

%%%%%%%%%%%%%%%%%%%%%%%%%%%%%%%%%%%%%%%%%%%%%%%%%%%%%%%%%%
\subsection{Phenomenological (IMRPhenom) models}
%%%%%%%%%%%%%%%%%%%%%%%%%%%%%%%%%%%%%%%%%%%%%%%%%%%%%%%%%
In the frequency domain, phenomenological IMR (IMRPhenom) 
models have been presented as another way 
to construct full GW signals 
by combining the analytical PN/BHP results with NR simulations. 

The IMRPhenom model basically consists of three parts: 
the PN inspiral (Ins) part, merger-ringdown (MR) part,
and intermediate (Int) part between the former two parts.
The amplitude $A_{\rm{IMR}}(f)$ 
and the GW phase $\Phi_{\rm{IMR}}(f)$ are written in the form of 
\bea
\Phi_{\rm{IMR}}(f) &=& \phi_{\rm{Ins}}(f) \, \theta_{f_{1\phi}}^{-} 
+  \theta_{f_{1\phi}}^{+} \, \phi_{\rm{Int}}(f) \,
\theta_{f_{2\phi}}^{-}
+ \theta_{f_{2\phi}}^{+} \, \phi_{\rm{MR}}(f) \,, \cr
A_{\rm{IMR}}(f)  &=& A_{\rm{Ins}}(f) \, \theta_{f_{\rm 1A}}^{-} 
+  \theta_{f_{\rm 1A}}^{+} \, A_{\rm{Int}}(f) \,
\theta_{f_{\rm 2A}}^{-}
+ \theta_{f_{\rm 2A}}^{+} \, A_{\rm{MR}}(f) \,,
\eea
where the function $\theta_{f_0}^{\pm}$ is defined by
\beq \theta_{f_0}^{\pm} = \frac{1}{2}\left\{1 \pm \theta(f-f_0)\right\} \,;
\quad 
\theta(f-f_0) = 
\left \{ \begin{array}{lllll}
 -1 \,,  & & f < f_0 \,, \\
 1 \,,   & & f \ge f_0 \,. \\
 \end{array}
 \right .
\eeq
Here, we basically pattern after the notation in Refs.~\cite{Husa:2015iqa,Pratten:2020fqn}, 
and we introduced the certainly prepared transition frequencies 
$f_{1\phi}$, $f_{2\phi}$, $f_{\rm 1A}$ and $f_{\rm 2A}$. 

Each individual component is parametrised and calibrated against NR simulations.
For example, the amplitude and phase models of the inspiral part are 
based on extensions of those of TaylorF2 models with calibration parameters. 
Their model functions in the state-of-the-art PhenomX framework  
take the form~\cite{Pratten:2020fqn}
\bea
A_{\rm Ins} &=& 
A_{\rm TF2}
+ 
\sqrt{\frac{2 \eta}{ 3 \pi^{1/3}}}\,f^{-7/6}\,
\sum_{i = 1}^{3}\, \rho_i (\pi f)^{(6+i)/3} \,, \cr
%%%%%%%%%%%%%%%%%%%%%%%%%%%%%%%%%%%%%%%%%%%%%
\phi_{\rm Ins} &=& 
\phi_{\rm TF2}
+ \frac{1}{\eta}
\left(\sigma_{0} + \sigma_{1} f
+ \frac{3}{4} \sigma_{2} f^{4/3}
+ \frac{3}{5} \sigma_{3} f^{5/3}
+ \frac{1}{2} \sigma_{4} f^{2} 
+ \frac{3}{7} \sigma_{5} f^{7/3} 
\right) \,,
\eea
where $A_{\rm TF2}$ and $\phi_{\rm TF2}$ are (essentially) the same 
as TaylarF2 models in Eqs.~\eqref{A-F2} and~\eqref{F2-inf}. 
The free parameters $\rho_i$ and $\sigma_j$ ($j = 0,\,1,\,2,\,3,\,4$ and $5$)
are phenomenological, pseudo-PN coefficients calibrated against NR data sets.

In Ref.~\cite{Ajith:2007kx}, the first frequency-domain IMR waveform, IMRPhenomA model, 
was presented for non-spinning BBHs.
The spinning, nonprecessing BBH waveform
is called the IMRPhenomB model~\cite{Ajith:2009bn},
and it is later improved to IMRPhenomC model~\cite{Santamaria:2010yb}; 
see also Section~II-b in Ref.~\cite{Abadie:2011kd} about IMRPhenomA/B models. 
In the classical GW data analysis of
the LIGO fifth science run (S5)~\cite{Abadie:2011kd},
the IMRPhenomA and IMRPhenomB models were used
as an IMR theoretical template; 
note, however, that IMRPhenomA/B/C models have been deprecated and 
they are no longer used in the modern LAL simulation. 
%(note that
%is also useful to briefly see the ).
All models in IMRPhenom family are available
at \url{https://lscsoft.docs.ligo.org/lalsuite/lalsimulation/group___l_a_l_sim_i_m_r_phenom__c.html}

%Ref.~\cite{lal_imr_phenom}.

\subsubsection{IMRPhenomD family}

%%%%%%%%%%%%%%%%%%%%%%%%%%%%%%%%%%%%%%%%%%%%%%%%%
{\it IMRPhenomD/HM}.---
%%%%%%%%%%%%%%%%%%%%%%%%%%%%%%%%%%%%%%%%%%%%%%%%%
In Refs.~\cite{Husa:2015iqa,Khan:2015jqa},
the IMRPhenomD model is presented for the dominant 
$(\ell,\,|m|) = (2,\,2)$ modes of spinning, nonprecessing binaries.
For the analysis of GW170817 in Ref.~\cite{Abbott:2018wiz}, 
using the IMRPhenomD model with the help of NRTidal for the tidal effects approximants~\cite{Dietrich:2018uni,Dietrich:2017aum,Dietrich:2019kaq}  
$$
{\rm IMRPhenomD\_NRTidal} = {\rm IMRPhenomD} + {\rm NRTidal}
$$
was used as a signal model of BNSs.
In Ref.~\cite{London:2017bcn}, based on the IMRPhenomD model,
the subdominant harmonic modes have been included in the waveform model,
giving origin to the IMRPhenomHM model.

%\subsubsection{IMRPhenomP/Pv2/Pv3/IMRPhenomPv3HM}
%%%%%%%%%%%%%%%%%%%%%%%%%%%%%%%%%%%%%%%%%%%%%%%%%
{\it IMRPhenomP/Pv2/Pv3/Pv3HM}.---
%%%%%%%%%%%%%%%%%%%%%%%%%%%%%%%%%%%%%%%%%%%%%%%%%
%For the IMRPhenomP/Pv2/Pv3 models, see Ref.~\cite{Khan:2018fmp} and references therein.
The IMRPhenomP~\cite{Hannam:2013oca} and IMRPhenomPv2 models~\cite{Husa:2015iqa} 
are for precessing binaries, based on the nonprecessing IMRPhenomC and IMRPhenomD models, respectively, 
and have the single precession spin 
(to rotate the nonprecessing signals in the co-precessing frame).
The improved IMRPhenomPv3 model~\cite{Khan:2018fmp} has two independent spins 
in the precession dynamics, making use of the results of the multi-timescale analysis 
of the (conservative) PN precession dynamics~\cite{Chatziioannou:2017tdw} 
to cover a broader region of the parameter space than that of the IMRPhenomPv2 model; recall the radiation-reaction and precession time scales in Eqs.~\eqref{t_RR} and~\eqref{t_P} 
that imply $t_{\rm p}/t_{\rm RR} \sim v^3 \ll 1$.
This model is further extended to IMRPhenomPv3HM model~\cite{Khan:2019kot} 
to include subdominant GW modes, which is based on the (nonprecessing) IMRPhenomHM model.

The IMRPhenomPv2 model was used in the detailed study on the properties of GW150914~\cite{TheLIGOScientific:2016wfe}.
In GWTC-1~\cite{LIGOScientific:2018mvr}, the IMRPhenomPv2 model was also used 
as a model for BBH coalescence.
The IMRPhenomPv3HM model was used for GW190521~\cite{Abbott:2020tfl} 
and GW190814~\cite{Abbott:2020khf} as well as Ref.~\cite{LIGOScientific:2020stg} 
and GWTC-2~\cite{Abbott:2020niy} to analyse potential BBH signals 
(we note that a fast and accurate NR surrogate model NRSur7dq4~\cite{Varma:2019csw} 
has also been used for GW190521 and GWTC-2).

%\subsubsection{IMRPhenomPv2\_NRTidal/NRTidalv2}
%%%%%%%%%%%%%%%%%%%%%%%%%%%%%%%%%%%%%%%%%%%%%%%%%
{\it IMRPhenomPv2\_NRTidal/NRTidalv2}.---
%%%%%%%%%%%%%%%%%%%%%%%%%%%%%%%%%%%%%%%%%%%%%%%%%
Including spin-precessing with the Pv2 style~\cite{Husa:2015iqa}
and tidal interactions with the NTRidal~\cite{Dietrich:2017aum},
we have
$$
{\rm IMRPhenomPv2\_NRTidal} = {\rm IMRPhenomPv2}
+ {\rm NRTidal}\,.
$$
This is called as the 
IMRPhenomPv2\_NRTidal (PhenomPv2NRT) model~\cite{Dietrich:2018uni}.
In Ref.~\cite{Dietrich:2019kaq}, 
an improved version of IMRPhenomPv2\_NRTidal
has been presented as the IMRPhenomPv2\_NRTidalv2 model.
In the detailed analysis of GW170817~\cite{Abbott:2018wiz},
this IMRPhenomPv2\_NRTidal model has been used 
as the reference model.
In GWTC-1~\cite{LIGOScientific:2018mvr}, 
this was used as a frequency-domain model for GW170817.
Also, to analyze GW190425~\cite{Abbott:2020uma} with total mass $\sim 3.4M_\odot$ 
and any sources in GWTC-2~\cite{Abbott:2020niy} that had evidence for at least one binary component 
below $3M_\odot$, this was used as the signal model.

%\subsubsection{IMRPhenomNSBH}
%%%%%%%%%%%%%%%%%%%%%%%%%%%%%%%%%%%%%%%%%%%%%%%%%
{\it IMRPhenomNSBH}.---
%%%%%%%%%%%%%%%%%%%%%%%%%%%%%%%%%%%%%%%%%%%%%%%%%
For spinning, nonprecessing NSBH binaries (with a non-spinning NS and a spinning BH), 
the IMRPhenomNSBH model has been developed~\cite{Thompson:2020nei}, 
based on the amplitude of the IMRPhenomC model and the phase of the IMRPhenomD\_NRTidalv2 model.
The IMRPhenomNSBH model was used for the analysis of GW190814~\cite{Abbott:2020khf} 
and potential NSBH sources in GWTC-2~\cite{Abbott:2020niy}.

%%%%%%%%%%%%%%%%%%%%%%%%%%%%%%%%%%%%%%%%%%%%
\subsubsection{IMRPhenomX family}
%%%%%%%%%%%%%%%%%%%%%%%%%%%%%%%%%%%%%%%%%%%%%%

IMRPhenomX family is an entirely new Phenom pipeline, 
superseding the IMRPhenomD family.
The main improvements from PhenomD include, e.g., (i) the larger number of input NR waveforms 
for calibrations increased from $19$ to $652$, broadening the coverage of the 
mass ratio from $1:18$ to $1:1000$ with the help of BBH merger simulations 
in the small mass-ratio limit produced 
by EOB-BHP approach~\cite{Harms:2014dqa,Harms:2015ixa,Harms:2016ctx}; 
(ii) the higher dimensionality the model parameter space 
enlarged from $2$ to $3$, using the symmetric mass ratio and two spin components 
(orthogonal to the orbital plane); further improvements are summarized in Sec.X of 
Ref.~\cite{Pratten:2020fqn}. These refinements resolve various shortcoming of PhenomD family  
and drastically improve the accuracy.

The baseline models for the dominant $(\ell,\,m) = (2,\,2)$ modes of spinning, 
nonprecessing binaries is called IMRPhenomXAS~\cite{Pratten:2020fqn}. 
This model is then generalized to IMRPhenomXHM model~\cite{Garcia-Quiros:2020qpx,Garcia-Quiros:2020qlt} 
to include subdominant harmonic modes of nonprecessing binaries, 
and further to the IMRPhenomXPHM model~\cite{Pratten:2020ceb} for precessing binaries with ``twisting-up'' the nonprecessing waveform 
using the (Pv3-style) double-spin approach developed in Ref.~\cite{Chatziioannou:2017tdw}. 
%see also e.g. Section~5.3 of Ref.~\cite{Hannam:2013pra} 
%about the (Pv2-style) single precession-angle approach.

All the models in IMRPhenomX family are available at 
\url{https://lscsoft.docs.ligo.org/lalsuite/lalsimulation/group___l_a_l_sim_i_m_r_phenom_x__c.html}; 
see Appendix C of Ref.~\cite{Garcia-Quiros:2020qpx} for the technical details of implementations, leading to significantly faster waveform production without compromising
on accuracy.

\subsubsection{IMRPhenomTP}
IMRPhenomTP Ref.~\cite{Estelles:2020osj} is a time-domain phenomenological model 
for the dominant $(\ell,\,m) = (2,\,2)$ modes of spinning precessing BBHs, 
making use of the "twisting up" approximation~\cite{Schmidt:2012rh} 
(see also, e.g., Section~5.3 of Ref.~\cite{Hannam:2013pra}) 
to the nonprecessing BBHs, based on TaylorT3 approximants.

%%%%%%%%%%%%%%%%%%%%%%%%%%%%%%%%%%%%%%%%%%%%%%%%%%%%%%%
\subsubsection{GIMR for modified theory of gravity}
%%%%%%%%%%%%%%%%%%%%%%%%%%%%%%%%%%%%%%%%%%%%%%%%%%%%%%%%

Until now we have (implicitly) assumed that the gravity theory 
is described by Einstein's GR. 
However, it is not the only relativistic theory of gravity. 
Indeed, motivated by the recent observation of accelerating expansion of the universe,  
there is a growing interest to consider alternative gravity theory other than GR.
Yet, each candidate gravity theory has to be experimentally verified, 
and the GW signals from the coalescence of a compact object binary 
allow a unique test of gravity theories in the strong curvature regime; 
this topic is covered in Chapter~41 by Yagi and Carson in this book 
with a lot more details.

Although we can consider model-dependent GW waveforms for each modified theory of gravity,
it is possible to formulate a model-independent waveform  
that phenomenologically captures the main features of a wider class of the modified theory of gravity.
For instance, the {\small G}IMR model is prepared by introducing deformations
to the phase of the frequency-domain IMRPhenom waveform model in GR (see also a parametrized post-Einsteinian framework in Ref.~\cite{Yunes:2009ke}).
The standard pipeline for model-independent {\small G}IMR waveforms is called TIGER (Test Infrastructure for GEneral Relativity)~\cite{Agathos:2013upa}; 
see Chapter~44 by Broeck in this book for a detailed description.

In Refs.~\cite{TheLIGOScientific:2016src} for GW150914, 
\cite{Abbott:2018lct} for GW170817, 
\cite{LIGOScientific:2019fpa} for GWTC-1,
and \cite{Abbott:2020jks} for GWTC-2, 
the {\small G}IMR model has been also
used to test the dipole radiation at $-1$PN order, which is absent in GR 
(recall our previous discussion on the quadrupole formalism for GW generation)
but is a common prediction of modified theory of gravity due to the existence 
of additional scalar degrees of freedom mediating long-range interactions.
See Chapter~40 by De Laurentis and De Martino in this book for more details.

%%%%%%%%%%%%%%%%%%%%%%%%%%%%%%%%%%%%%%%%%%%%%%%%%%%%%%%%%%%%%%
\section{Conclusion}
%%%%%%%%%%%%%%%%%%%%%%%%%%%%%%%%%%%%%%%%%%%%%%%%%%%%%%%%%%%%%%

The development of the theoretical GW templates of coalescing 
compact object binaries was initiated within the PN approach, focusing on the adiabatic inspiral phase.  
However, after the 2005 breakthroughs in NR simulations of 
BBHs~\cite{Pretorius:2005gq,Campanelli:2005dd,Baker:2005vv}
and the observations of GW events by LIGO and Virgo, 
learning that almost all the GW signals detected so far 
have both merger and ringdown in the sensitive frequency band of LIGO, Virgo and KAGRA, 
complete models with inspiral, merger, and ring-down phases are under vigorous development.

The theoretical construction of GW waveforms
for the entire coalescence makes full use of known 
analytical approximation scheme to other methods than 
the PN approximation to GR, e.g. the BHP (and self-force) theory 
%(see, e.g., Chapter 36 by Pound and Wardell in this book)
as well as cutting-edge NR simulations. 
%(see, e.g., Chapter 34 by Zhao et al. in this book). 

This motivation has continuously driven a concerted effort 
by GW theorists and data analysists to develop accurate and efficient  
waveforms of compact-object binary mergers~\cite{lalsuite}, 
and we have provided a broad (but yet small-corner) overview of this active subject. 
We conclude our chapter by listing some of open challenges and prospects.

\begin{itemize}
\item
Although we have exclusively discussed quasicircular binaries in this chapter, 
the observation of nonzero eccentricity will be a smoking gun of the formation scenario 
of binary systems: evolution of isolated binaries 
(e.g., circularized by binary interactions and GW radiation; recall Eq.~\eqref{dote})
versus dynamically formed binaries in dense stellar environments
(e.g., the Kozai-Lidov mechanism~\cite{Kozai:1962zz,1962P&SS....9..719L}).
\item
The ground-based GW detectors, LIGO, Virgo, and KAGRA, have a peak of sensitivity around 100\,Hz.
For future plans of ground-based detectors, e.g., 
KAGRA+~\cite{Michimura:2019cvl}, Voyager (\url{https://dcc.ligo.org/LIGO-G1602258/public}), 
Einstein Telescope~\cite{Hild:2008ng} and Cosmic Explorer~\cite{Reitze:2019iox},
significant sensitivity improvements are expected both in low- and high-frequency bands.
Enhanced low-frequency part would stretch the visible range 
of BBHs and NS-BH inspirals to heavier masses. 
%that otherwise would merge before entering the detector's sensitivity band.
At the same time, improvements in the high frequency part would enable us 
to observe the BNS merger phase more accurately, 
where the (equation-of-state dependent) finite-size effects of NSs 
become particularly pronounced. 
\item
The observation of GW190814 showed a binary system with mass ratio
around 10:1.
The remnant compact object of GW190521~\cite{Abbott:2020tfl}
has been considered as an intermediate mass BH.
The combination of these two events may suggest the existence of intermediate
mass black holes and the possibility to observe in the future binary
systems with an individual mass ratio around $1:100$.
In LALSimulation~\cite{lalsuite},
a program for ``extreme'' mass ratios is present
at \url{https://lscsoft.docs.ligo.org/lalsuite/lalsimulation/group___l_a_l_sim_i_m_r_phenom_x__c.html},
%in Ref.~\cite{lal_imr_phenom_x}
%It is calibrated to mass ratios from 1 to 1000.
calibrated to NR waveforms for mass ratio from $q=1$ to $20$, 
and the region between $q = 20$ and $1000$, where no NR waveforms are available,  
is covered by the waveform of extreme mass-ratio inspirals (EMRIs) 
(based on the BHP theory with the EOB orbital dynamics in the small mass-ratio limit
~\cite{Harms:2014dqa,Harms:2015ixa,Harms:2016ctx}), 
with the mass ratio region $200<q<1000$ covered by 
the extrapolation based on these waveforms.
Further synergy with EMRI waveforms is ongoing: 
see, e.g., Refs.~\cite{Berry:2019wgg,Rifat:2019ltp,vandeMeent:2020xgc}
%Refs.~\cite{AmaroSeoane:2007aw,Barack:2009ux,Berry:2019wgg}
and references therein.
\item
The planned space-based GW observatory, e.g., 
LISA~\cite{Audley:2017drz}, (B-)DECIGO~\cite{Kawamura:2020pcg,Seto:2001qf,Nakamura:2016hna}, 
and TianQin~\cite{Mei:2020lrl} will observe much longer-length inspiral signals 
of compact object binaries than LIGO, Virgo and KAGRA. 
They will allow us to measure the binary parameters with exquisite precision, 
particularly in the context of the multiband GW astronomy~\cite{Sesana:2016ljz,Vitale:2016rfr,Isoyama:2018rjb}. 
At the same time, however, the benefit of such an observation is gained 
only when one is able to construct much more accurate templates for the inspiral phase
than currently available models: 
their phase coherence has to be maintained over $O(10^6)$ GW cycles 
(and one should also devise a consistent data analysis technique 
to process such a long-length GW signals). 
\item Our presentation does not cover the \textit{efficiency} aspects 
in the waveform modelling. The matched-filtering search of GW signal needs 
a (so named) ``template bank'' by nature (see, e.g., Chapter~43 by Krolak in this book), 
designed to efficiently cover a parameter space as large as possible. 
This is a computationally expensive and challenging task, demanding different investigations.  
A common strategy is to use ``effective'' parameters 
such as the chirp mass $\cal M$~\eqref{def-chirpM}, 
the reduced spin $\chi_{\rm PN}$~\eqref{def-chiPN}, 
and the binary tidal deformability ${\tilde \Lambda}$~\eqref{def-tildeLambda}, 
to reduce the numbers of the dimension in the parameter space by focusing
on those combinations of astrophysical parameters which
affect most prominently the waveform.
An alternative strategy is a reduced-order modeling (ROM)~\cite{Purrer:2014fza} 
and a surrogate model (for NR waveforms)~\cite{Varma:2019csw}.
\end{itemize}

%%%%%%%%%%%%%%%%%%%%%%%%%%%%%%%%%%%%%%%%%%%%%%%%%%
\section{Acknowledgments}
%%%%%%%%%%%%%%%%%%%%%%%%%%%%%%%%%%%%%%%%%%%%%%%%%%

The authors warmly thank Maria Haney for reading 
the manuscript and improving it with her suggestions.
S.~I. acknowledges support from STFC through Grant No. ST/R00045X/1. 
S.~I. also thanks the financial support from the Ministry of Education, MEC, 
during his stay at IIP-Natal-Brazil 
and acknowledges networking support by the GWverse COST Action CA16104, 
“Black holes, gravitational waves and fundamental physics."
The work of R.~S. is partially supported by CNPq. 
H.~N. acknowledges support from JSPS KAKENHI Grant Nos. JP16K05347 and JP17H06358.

\bibliographystyle{spbasic}

\bibliography{references}

%%%%%%%%%%%%%%%%%%%%%%%%%%%%%%%%%%%%%%%%%%%%%%%%%%
\end{document}